\documentclass[aps,prc,groupedaddress,showpacs,nofootinbib]{revtex4}

\usepackage[dvips]{graphicx}

\newcommand{\be}{\begin{equation}}
\newcommand{\ee}{\end{equation}}
\newcommand{\bea}{\begin{eqnarray}}
\newcommand{\eea}{\end{eqnarray}}
\newcommand{\myunit}[1]{\; \text{#1} }
\newcommand{\myvec}[1]{\,\vec{#1}\,}
\newcommand{\reffig}[1]{Fig.~\ref{#1}}
\newcommand{\refeq}[1]{Eq.~(\ref{#1})}
\newcommand{\refch}[1]{Sec.~\ref{#1}}
\newcommand{\refcite}[1]{Ref.~\cite{#1}}
\newcommand{\refscite}[1]{Refs.~\cite{#1}}
\newcommand{\refetal}[1]{\emph{et~al.}~\cite{#1}}
\newcommand{\slashp}{p\hspace{-1.5mm}/}
\newcommand{\slashq}{q\hspace{-1.7mm}/}

\begin{document}

\title{Neutral current neutrino-nucleus interactions at intermediate energies}

\author{T.~Leitner}
\email[Electronic address: ]{tina.j.leitner@theo.physik.uni-giessen.de}
\author{L.~Alvarez-Ruso}
\author{U.~Mosel}
\affiliation{Institut f\"ur Theoretische Physik, Universit\"at Giessen, Germany}

\date{June 28, 2006}

\begin{abstract}

We have extended our model for charged current neutrino-nucleus interactions 
developed in Phys.~Rev.~C~\textbf{73}, 065502 (2006) to neutral current reactions. For the elementary neutrino-nucleon interaction, we take into account quasielastic scattering, $\Delta$ excitation and the excitation of the resonances in the second resonance region. Our model for the neutrino-nucleus collisions includes in-medium effects such as Fermi motion, Pauli blocking, nuclear binding, and final-state interactions. 
They are implemented by means of the Giessen Boltzmann-Uehling-Uhlenbeck (GiBUU) coupled-channel transport model.
This allows us to study exclusive channels, namely pion production and nucleon knockout. We find that final-state inter\-actions modify considerably the distributions through rescattering, charge-exchange, and absorption. Side-feeding induced by charge-exchange scattering is important in both cases. In the case of pions, there is a strong absorption associated with the in-medium pionless decay modes of the $\Delta$, while nucleon knockout exhibits a considerable enhancement of low energy nucleons due to rescattering. At neutrino energies above $1$ GeV, we also obtain that the contribution to nucleon knockout from $\Delta$ excitation is comparable to that from quasielastic scattering. 

\end{abstract}

\pacs{13.15.+g, 25.30.Pt, 25.30.-c, 23.40.Bw, 24.10.Lx, 24.10.Jv}

\maketitle

\section{Introduction}

Neutrino interactions are classified as charged current (CC) or neutral current (NC) processes depending on whether a $W$ or a $Z$ boson is exchanged. In the first case, a charged lepton is emitted, whereas the neutrino preserves its nature in the second one. The existence and nature of neutral currents played an important role in the establishment of the Standard Model of electroweak interactions. Unfortunately, the experimental study of NC neutrino interactions is a demanding task due to the considerable difficulties of collecting data on reactions with cross sections even smaller than those of CC processes, and in which the outgoing neutrino leaves no signal, so that the event identification has to rely on the detection of one or more hadrons. Still, measurements of NC quasielastic (QE) scattering and pion production were performed at BNL~\cite{Ahrens:1986xe}, ANL~\cite{Barish:1974fe,Derrick:1980nr} and Gargamelle~\cite{Krenz:1977sw,Pohl:1978iy} using deuterium or heavier targets such as carbon, aluminum or a propane-freon mixture.       
 
Nowadays, the discovery of neutrino oscillations has renewed the interest in a better determination of the neutrino-nucleus cross sections, aimed to achieve a better understanding of neutrino fluxes and background processes at current and future experiments. In particular, it is well known that NC $\pi^0$ production in the forward direction is a relevant source of background for $\nu_e$ appearance experiments. In fact, both K2K~\cite{Nakayama:2004dp} and MiniBooNE~\cite{Raaf:2004ty} have recently measured this reaction and the proposed MINER$\nu$A experiment plans to do it in the future~\cite{Harris:2004iq}. At $E_\nu < 2$~GeV, pion production proceeds mainly through resonance excitation, predominantly of the $\Delta (1232)$ resonance~\cite{Fogli:1979qj,Rein:1980wg}. But when pions are produced in the nucleus, their final-state interaction (FSI) with the nucleons (elastic and charge-exchange scattering, absorption) is an essential ingredient of any realistic theoretical description of this reaction. In Ref.~\cite{Paschos:2000be}, NC pion production in nuclei was investigated. In that model, pions, initially produced via $\Delta(1232)$, $P_{11}(1440)$, and $S_{11}(1532)$ excitation, undergo a random walk through the nucleus where the pions can change directions but not energy. 

NC neutrino-nucleus interactions are also relevant to answer a fundamental question of hadronic structure, namely, the strange-quark contribution to the nucleon spin. Purely isovector CC processes do not depend on the strange form factors. Since a non-zero strange axial form factor changes the NC QE cross section on protons and neutrons in different ways (see \refch{sec:qe} for more details), the ratio $R(p/n)$ of these two cross sections is very sensitive to the strange spin, as pointed out by Garvey \emph{et al.}~\cite{Garvey:1992qp}. Due to the technical difficulties of neutron detection, the proposed FINeSSE experiment plans to measure the neutral to charged current ratio $R$(NC/CC). In any case, the study of those ratios involves both neutrons and protons so it must be performed using nuclear targets. This fact has stimulated a considerable amount of theoretical work aiming at the description of nuclear effects in NC nucleon knockout reactions. The relativistic Fermi gas description of the nucleus has been adopted in many calculations~\cite{Horowitz:1993rj,Barbaro:1996vd,Alberico:1997vh,Alberico:1997rm}; others use wave functions for the bound nucleons, obtained in relativistic mean-field models~\cite{Meucci:2004ip,Martinez:2005xe,vanderVentel:2005ke,Meucci:2006ir}. The shell model~\cite{Alberico:1997vh,Alberico:1997rm} and the continuum random-phase approximation~\cite{Kolbe:1994xb,Jachowicz:1998fn} have also been applied. The effect of mesons exchange currents has been evaluated by Umino and collaborators~\cite{Umino:1994wu}. Furthermore, the input from scaling analysis of electron scattering data has been used to obtain NC cross sections~\cite{Amaro:2006pr}. Final-state interactions of the knocked out nucleons is neglected in several studies~\cite{Horowitz:1993rj,Barbaro:1996vd,vanderVentel:2005ke} while many treat them with the distorted wave impulse approximation~\cite{Alberico:1997vh,Alberico:1997rm,Meucci:2004ip,Martinez:2005xe,Meucci:2006ir} and with a multiple scattering Glauber model at higher energies~\cite{Martinez:2005xe}. The problem is that in those approaches it is not possible to take into account nucleon rescattering leading to energy losses, charge exchange and multiple nucleon emissions. This can be achieved in Monte Carlo models~\cite{Nieves:2005rq}. It is widely accepted that nuclear effects cancel for the ratios of cross sections. However, this is not the case for side-feeding effects caused by charge exchange scattering, which are not negligible if the elementary cross section on protons and neutrons differ. Such changes in the ratios caused by FSI were obtained by Nieves \emph{et al.}~\cite{Nieves:2005rq} and we will confirm them here. All these theoretical models consider only QE processes, but one should bear in mind that for neutrinos of $\sim 1$~GeV, the excitation of resonances also contributes to nucleon knockout; those resonance excitation events where a nucleon is emitted but there are no pions in the final state (or they are produced but not detected) represent a source of background for the strange axial form factor measurements, which should be well understood.

In a recent article~\cite{Leitner:2006ww} we have studied CC neutrino-nucleus interactions in the region of the QE and $\Delta$ peaks. Here we extend the model to the NC case. There are three main ingredients in our model: elementary processes, in-medium modifications and FSI. QE scattering and $\Delta$ excitation on the nucleon are treated in a relativistic formalism, incorporating state-of-the-art parameterizations of the form factors for both the nucleon and the $N-\Delta$ transition. It is, however, known that the excitation of heavier resonances is not negligible above $E_\nu \approx 1.5$~GeV~\cite{Fogli:1979qj,Rein:1980wg}. Therefore, in this article, we also consider the excitation of the $N^*$ states $P_{11}(1440)$, $D_{13}(1520)$ and $S_{11}(1535)$, by means of the Rein and Sehgal model~\cite{Rein:1980wg}, which has been extensively used in the simulations of neutrino experiments~\cite{Casper:2002sd,Gallagher:2002sf,Hayato:2002sd}.
We should, however, recall that new information on the $N-N^*$ electromagnetic transition form factors is available from the analysis~\cite{Tiator:2003uu} of recent electron scattering data. Hence, there is room for improvement in this part of the model~\cite{Lalakulich:2006sw}. Next, we take into account nuclear effects: Fermi motion, Pauli blocking and the binding of the nucleons in a density and momentum dependent mean-field. Finally, FSI are implemented in the framework of the semiclassical coupled-channel transport theory: the Giessen BUU model. With these tools we investigate pion production and nucleon knockout in neutral current neutrino-nucleus interactions at intermediate energies.

Next, we present our model for the elementary cross sections emphasizing the structure of the neutral hadronic currents and the role of strange form factors. Following an overview of the nuclear model and final-state interactions, we present our results and compare these to other calculations. Summary and conclusions are given at the end.

\section{Neutrino-nucleon reactions}

In this section we present the model we have adopted for the description of elementary neutrino-nucleon interactions, emphasizing the aspects that are specific of neutral current processes. Further details on our corresponding approach to charged current processes can be found in \refcite{Leitner:2006ww}.

We consider neutral current reactions of the type,
\be
\nu(k) + N(p) \to \nu(k') + X(p'), 
\ee
with $k_{\alpha}=\left(E_{\nu},\myvec{k}\right)$, $k'_{\alpha}=\left(E'_\nu,\myvec{k\,'}\right)$, $p_{\alpha}=\left(E,\myvec{p}\right)$ and $p'_{\alpha}=\left(E',\myvec{p\,'}\right)$. 
The cross section can be cast as an integral over the azimuthal angle of the outgoing neutrino
\be
\frac{{\rm d}^2\sigma_{\nu N}}{{\rm d}Q^2{\rm d}E'_{\nu}} = \int {\rm d} \phi \; \frac{1}{64 \pi^2} \frac{1}{| k \cdot p|} \frac{1}{E_{\nu}} \delta\left({p'}^2-{M'}^2\right) |\bar{\mathcal{M}}|^2 \,,  \label{eq:crosssection}
\ee
where $Q^2=-(k-k')^2$ and $M'$ is the (invariant) mass of the outgoing baryon.
For neutral current (NC) interactions, the matrix element squared, summed and averaged over spins,  
\be
|\bar{\mathcal{M}}|^2=\frac{G_F^2}{2} L_{\alpha \beta} H^{\alpha \beta},  \label{eq:matrixel}
\ee
is given in terms of the  Fermi constant $G_F=1.16637 \times 10^{-5} \myunit{GeV}^{-2}$, the leptonic tensor  $L_{\alpha \beta}$ and the hadronic tensor $H^{\alpha \beta}$. While the calculation of $L_{\alpha \beta}$ is straightforward, the hadronic current entering $H^{\alpha \beta}$ has to be parametrized in terms of form factors and thus depends on the specific reaction. Two processes dominate the reaction at neutrino energies up to about 1.5 GeV, namely quasielastic scattering,
\be
\nu  n \to  \nu  n, \quad \quad   \nu  p \to  \nu  p, 
\ee
and $\Delta (1232) P_{33}$ excitation,
\be
\nu  n \to  \nu  \Delta^0, \quad \quad   \nu  p \to  \nu  \Delta^+. 
\ee
Here, in addition, we consider the excitation of the $N^*$ states $R= P_{11}(1440)$, $D_{13}(1520)$ and $S_{11}(1535)$, which form the so-called second resonance region,
\be
\nu  n \to  \nu  R^0, \quad \quad   \nu  p \to  \nu  R^+. 
\ee

\subsection{Quasielastic scattering \label{sec:qe}} 

The hadronic tensor $H^{\alpha \beta}$ for quasielastic scattering is determined by the hadronic current $J_{\alpha}^{QE}$ given by
\bea
J_{\alpha}^{QE}&=& \langle N' |J_{\alpha}^{NC}(0) | N \rangle  \nonumber \\ [0.2cm]
               &=& \langle N' |(V_{\alpha}^{NC} - A_{\alpha}^{NC})(0) | N \rangle  \nonumber \\ [0.2cm]
               &=&  \bar{u}(p') B_{\alpha} u(p)
\eea
with $q_{\alpha}=p'_{\alpha}-p_{\alpha}$, $N=p,n$ and 
\be
B_{\alpha } =\left(\gamma_{\alpha} - \frac{\slashq \,q_{\alpha}}{q^2} \right) \tilde{F}_1^N + \frac{i}{2 M_N} \sigma_{\alpha \beta} q^{\beta} \tilde{F}_2^N + \gamma_{\alpha} \gamma_5 \tilde{F}_A^N +  \frac{q_{\alpha}}{M_N}\gamma_5 \tilde{F}_P^N, 
\ee
where $M_N$ denotes the nucleon mass; $\tilde{F}_{1,2}^N$, $\tilde{F}^N_A$ and  $\tilde{F}^N_P$ are the vector, axial and pseudoscalar form factors, respectively. Due to time invariance, these form factors are real functions of $Q^2$. In the expression for the cross section, $\tilde{F}^N_P$ appears only multiplied by the neutrino mass so we ignore it from now on. The term $(\slashq \,q_{\alpha})/q^2$ ensures vector current conservation even if the masses of the initial and final nucleons differ, as might be the case in the presence of a momentum-dependent nuclear mean-field potential.
The hadronic tensor $H^{\alpha \beta}$ then follows as
\be
H^{\alpha \beta}_{QE}=\frac12 {\rm Tr} \left[ (\slashp \, + M) \tilde{B}^{\alpha} (\slashp' \, + M') B^{\beta}  \right] \label{eq:hadrtensorQE}
\ee
with
\be
\tilde{B}_{\alpha}=\gamma_0 B_{\alpha}^{\dagger} \gamma_0
\ee
and --- in the free nucleon case --- with $M=M'=M_N$.

The vector current has the form
\be
V_{\alpha}^{NC} =(1-2 \sin^2 \theta_W) V_{\alpha}^3 - 2 \sin^2 \theta_W \frac{1}{2} J_{\alpha}^Y - \frac{1}{2} J_{\alpha}^{s}\,,  
\ee
where $\theta_W$ is the weak-mixing angle ($\sin^2 \theta_W=0.2228$), $V_{\alpha}^3$ is the third component of the isovector current, $J_{\alpha}^Y$ is the isoscalar (hypercharge) one and $J_{\alpha}^{s}$ stands for the strange part. All of these terms have the same Dirac structure. Therefore, the NC form factor can be written as    
\bea
\tilde{F}_{1,2}^p&=&\left(\frac12-2 \sin^2 \theta_W\right)F_{1,2}^p  - \frac12 F_{1,2}^n - \frac12 F_{1,2}^{s}, \\
\tilde{F}_{1,2}^n&=&\left(\frac12-2 \sin^2 \theta_W\right)F_{1,2}^n  - \frac12 F_{1,2}^p - \frac12 F_{1,2}^{s} 
\eea
in terms of the the standard Dirac and Pauli form factors of the nucleon $F^{p,n}_1$ and $F^{p,n}_2$ and a strange component $F_{1,2}^{s}$.

Analogously, the axial current consists of the third component of the isovector axial current and a strangeness part
\be
A_{\alpha}^{NC}=A_{\alpha}^3+\frac{1}{2} A_{\alpha}^{s}. 
\ee
This implies that 
\be
\tilde{F}_{A}^{p,n}=\pm  \frac12 F_{A}+ \frac12 F_{A}^{s},   \label{eq:fa}
\ee
where $F_{A}$ is the axial form factor for charged current QE scattering. Notice that here it appears with different signs for protons and neutrons. $F_{A}^{s}$ stands for the strange axial form factor.     

As in our earlier work~\cite{Leitner:2006ww}, we use the BBA-2003 parametrization~\cite{Budd:2003wb} for the non-strange vector form factors and a dipole ansatz with $M_A=1.0$~GeV for $F_A$ (See Ref.~\cite{Bernard:2001rs} for an overview on the extraction of $M_A$ from neutrino-nucleon scattering, pion electroproduction and muon capture).

The strangeness content of the nucleon, encoded in  $F_{1,2}^{s}$ and $F_{A}^{s}$, is still an open question. It can be investigated in a combined study of parity-violating polarized electron scattering and neutral current neutrino scattering. Parity-violating electron scattering is very sensitive to the strange electric and magnetic form factors (i.~e., to the strange vector form factors) and much less to the strange axial vector form factor (cf.~e.~g.~\refcite{Weise:2001}). The opposite holds for NC neutrino scattering. An extensive program on parity violation has evolved in the last years: the SAMPLE experiment at MIT/Bates~\cite{Spayde:2003nr}, HAPPEX~\cite{Aniol:2000at,Aniol:2005zg} and G0~\cite{Armstrong:2005hs} at JLab, and 
PVA4~\cite{Maas:2004ta} in Mainz have extracted linear combinations of the  strange electric and magnetic form factors at different $Q^2$ values. Recently, the strange electric and magnetic form factors were extracted from a combined set of available parity-violating electron scattering data by Young \refetal{Young:2006jc}. 
However, data on NC neutrino scattering, needed to determine the strange axial form factor, are scarce. The best measurement to date is the E734 experiment at BNL~\cite{Ahrens:1986xe}. It measured neutrino-proton and antineutrino-proton elastic scattering albeit with large systematical errors and only small statistics. Former attempts to extract the strange axial form factor from the data~\cite{Garvey:1992cg, Alberico:1998qw} faced the fact that, as pointed out by Alberico \refetal{Alberico:1998qw}, "the experimental uncertainty is still too large to be conclusive about specific values of the strange form factors of the nucleon" and a "rather wide range of values for the strange parameters is compatible with the BNL E734 data." The advent of new polarized electron scattering data from the above mentioned experiments changes appreciably the situation because, as shown by Pate~\cite{Pate:2003rk, Pate:2005ft}, it allows to perform a simultaneous determination of all (electric, magnetic and axial) strange form factors with small error bars, in spite of the uncertainties of the E734 data~\cite{Pate:2005bk}. But this is only possible in the region of $0.45 < Q^2 < 1.05$~GeV$^2$ where the E734 differential cross sections were measured, so new NC neutrino scattering data at low $Q^2$ are needed for a reliable extrapolation down to $Q^2=0$. The proposed FINeSSE experiment~\cite{finesseprop} will hopefully fill this gap.      

In view of the low sensitivity of NC neutrino scattering to the strange vector form factors and the large sensitivity to the axial one (see Ref.~\cite{leitner_diplom2} for a detailed study using the parameterizations of Garvey \refetal{Garvey:1992cg}) here, for the sake of simplicity, we choose to take
\bea
F_1^{s}(0)&=& 0, \\
F_2^{s}(0)&=& 0, 
\eea
and
\be
F_A^{s}(Q^2)=\frac{\Delta s}{\left(1+\frac{Q^2}{M_A^2}\right)^2} \,,
\ee
assuming that the strange axial mass is equal to the non-strange one. Here $\Delta s$ denotes the strange contribution to the nucleon spin. In line with \refcite{Alberico:1997vh}, we use $\Delta s= - 0.15$ and $\Delta s= 0$ as representative values.

\begin{figure*}
\includegraphics[width=16cm]{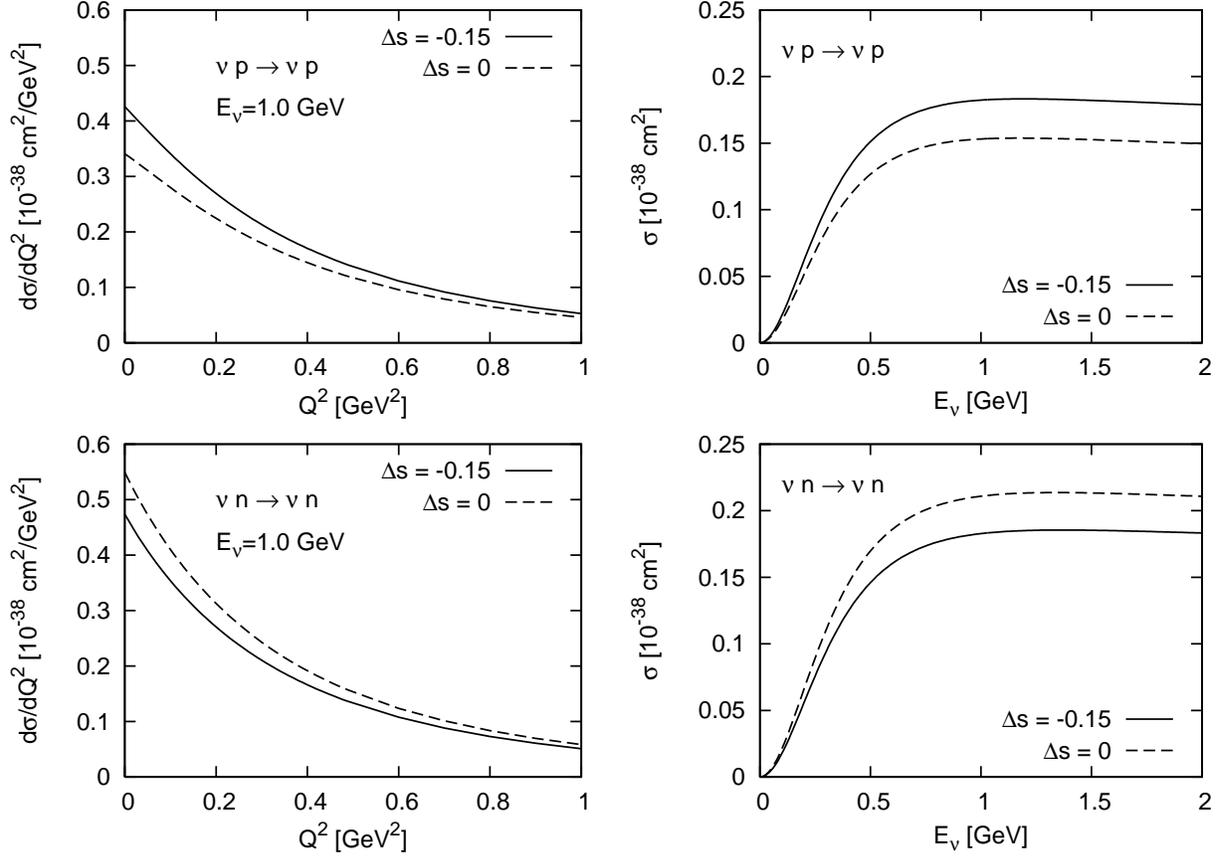}
\caption{Differential and integrated cross section for NC quasielastic scattering on protons (top) and neutrons (bottom). The solid lines denote the results with $\Delta s = -0.15$, the dashed lines denote the ones without the strange axial form factor ($\Delta s = 0$). \label{fig:nuQE}}
\end{figure*}

The cross sections for NC QE scattering on proton and neutron are shown in \reffig{fig:nuQE}. Note how the strange spin causes opposite effects on the cross sections for protons and neutrons; this is a direct consequence of the two different signs in \refeq{eq:fa}.

\subsection{Production of the $\Delta$ and of higher resonances} 

The hadronic tensor $H^{\alpha \beta}$ for $\Delta$ excitation is determined by the hadronic current given by~\cite{LlewellynSmith:1971zm} 
\bea
J_{\alpha}^{\Delta}&=& \langle \Delta^{0} |J_{\alpha}^{NC}(0) | n \rangle  \nonumber \\
&=& \langle \Delta^{+} |J_{\alpha}^{NC}(0) | p \rangle  \nonumber \\
          &=&  \bar{\psi}^{\beta}(p') B_{\beta \alpha } u(p)
\eea
with
\begin{widetext}
\bea
B_{\beta \alpha } &=& \left[ \frac{\tilde{C}_3^V}{M_N} (g_{\alpha \beta} \slashq \, - q_{\beta} \gamma_{\alpha})+
  \frac{\tilde{C}_4^V}{M_N^2} (g_{\alpha \beta} q\cdot p' - q_{\beta} p'_{\alpha}) 
  + \frac{\tilde{C}_5^V}{M_N^2} (g_{\alpha \beta} q\cdot p - q_{\beta} p_{\alpha}) + g_{\alpha \beta} \tilde{C}_6^V\right] \gamma_{5} \nonumber  \\  
  &&\; \;\;\;\;+ \frac{\tilde{C}_3^A}{M_N} (g_{\alpha \beta} \slashq \, - q_{\beta} \gamma_{\alpha})+
  \frac{\tilde{C}_4^A}{M_N^2} (g_{\alpha \beta} q\cdot p' - q_{\beta} p'_{\alpha})+
 {\tilde{C}_5^A} g_{\alpha \beta}
  + \frac{\tilde{C}_6^A}{M_N^2} q_{\beta} q_{\alpha},
\eea
\end{widetext}
where $\bar{\psi}^{\beta}(p')$ is the Rarita-Schwinger spinor for the $\Delta$, and $u(p)$ is the Dirac spinor for the nucleon. 
The hadronic tensor follows to
\be
H^{\alpha \beta}_{\Delta}=\frac12 {\rm Tr}\left[ (\slashp \, + M) \tilde{B}^{\alpha \rho} \Lambda_{\rho \sigma} B^{\sigma \beta}  \right]  \label{eq:hadrtensorDELTA}
\ee
with 
\be
\tilde{B}_{\alpha\beta}=\gamma_0 B_{\alpha\beta}^{\dagger} \gamma_0
\ee
and, for free nucleons, $M=M_N$.
The spin $3/2$ projection operator is given by 
\be
\Lambda_{\rho \sigma}=- \left(\slashp' \, + \sqrt{{p'}^2} \right) \left( g_{\rho \sigma} - \frac{2}{3} \frac{p'_{\rho } p'_{\sigma }}{{p'}^2} 
     + \frac{1}{3} \frac{p'_{\rho } \gamma_{\sigma} - p'_{\sigma } \gamma_{\rho}}{\sqrt{{p'}^2}} - \frac{1}{3} \gamma_{\rho} \gamma_{\sigma} \right).
\ee

In contrast to quasielastic scattering, where the neutral current is sensitive to the isoscalar quark content of the nucleon, the $N-\Delta$ transition is purely isovector. Therefore, the neutral current reduces to 
\be
J_{\alpha}^{NC}= (1-2 \sin^2 \theta_W) V_{\alpha}^3-A_{\alpha}^{3} \,,
\ee
where $V_{\alpha}^3$ and $A_{\alpha}^{3}$ are the third components of the vector and axial isovector currents, respectively. Thus, we have
\bea
\tilde{C}^V_i &=& (1-2 \sin^2 \theta_W) C^V_i, \\
\tilde{C}^A_i &=& C^A_i,
\eea
where $C^V_i$ and $C^A_i$ are the vector and axial charged current transition form factors for which we use the parametrization of Ref.~\cite{Paschos:2003qr} as in our earlier work \cite{Leitner:2006ww}.

The phenomenological information about the neutrino induced $N-N^*$ transition is far more scarce. Several articles have considered $N-N^*$ vector form factors derived from helicity amplitudes~\cite{Fogli:1979qj, Fogli:1979cz,Alvarez-Ruso:1997jr,Alvarez-Ruso:2003gj}, extracted from electron scattering experiments or calculated with different quark models~\cite{Rein:1980wg}, especially for the $P_{11}(1440)$~\cite{Li:1991yb, Cano:1998wz, Alberto:2001fy, Cardarelli:1996vn, Dong:1999cz}. In the context of neutrino scattering, the most recent study based on new JLAB electron scattering data has been performed by Lalakulich \refetal{Lalakulich:2006sw}. Experimental information on the $N-N^*$ axial form factors is very limited. Goldberger-Treiman relations have been derived for the axial couplings \cite{Fogli:1979qj, Fogli:1979cz}, but there is no information about the $Q^2$ dependence.
For the production of higher resonances, we use the matrix elements derived from the model of Rein and Sehgal~\cite{Rein:1980wg} who apply a quark model to calculate the vector and axial $N-N^*$ transitions.\footnote{Our matrix element $\mathcal{M}$ corresponds to $T$ in the notation of Rein and Sehgal \cite{Rein:1980wg}.} This model has been widely used in the simulation and analysis of many neutrino experiments~\cite{Casper:2002sd,Gallagher:2002sf,Hayato:2002sd}.
The $Q^2$ dependence of their form factors has the form of a modified dipole (see their Eq.~(3.12)). For the dipole masses, we use $M_A=1.032$~GeV and $M_V=0.84$~GeV (cf.~e.~g.~\refcite{Zeller:2003ey}). In this way we include the production of the resonances $P_{11}(1440)$, $D_{13}(1520)$ and $S_{11}(1535)$.

The finite width of the resonances is accounted for in the cross section of \refeq{eq:crosssection} by replacing  
\be
\delta\left({p'}^2-{M'}^2\right) \to  - \frac{1}{\pi} \mathcal{I}m \left(\frac{1}{{p'}^2-{M'}^2+i \sqrt{{p'}^2} \Gamma}\right),
\ee
where $M'$ is the pole mass of the resonance and $\Gamma$ is the energy dependent full decay width in the vacuum. 
We use a consistent set of resonance parameters taken from a single analysis, namely the one of Manley and Saleski~\cite{manley}. All the relevant decay channels as $\pi N$, $\pi \Delta$, $\rho N$ and $\eta N$ are included.

\begin{figure*}
\includegraphics[width=16cm]{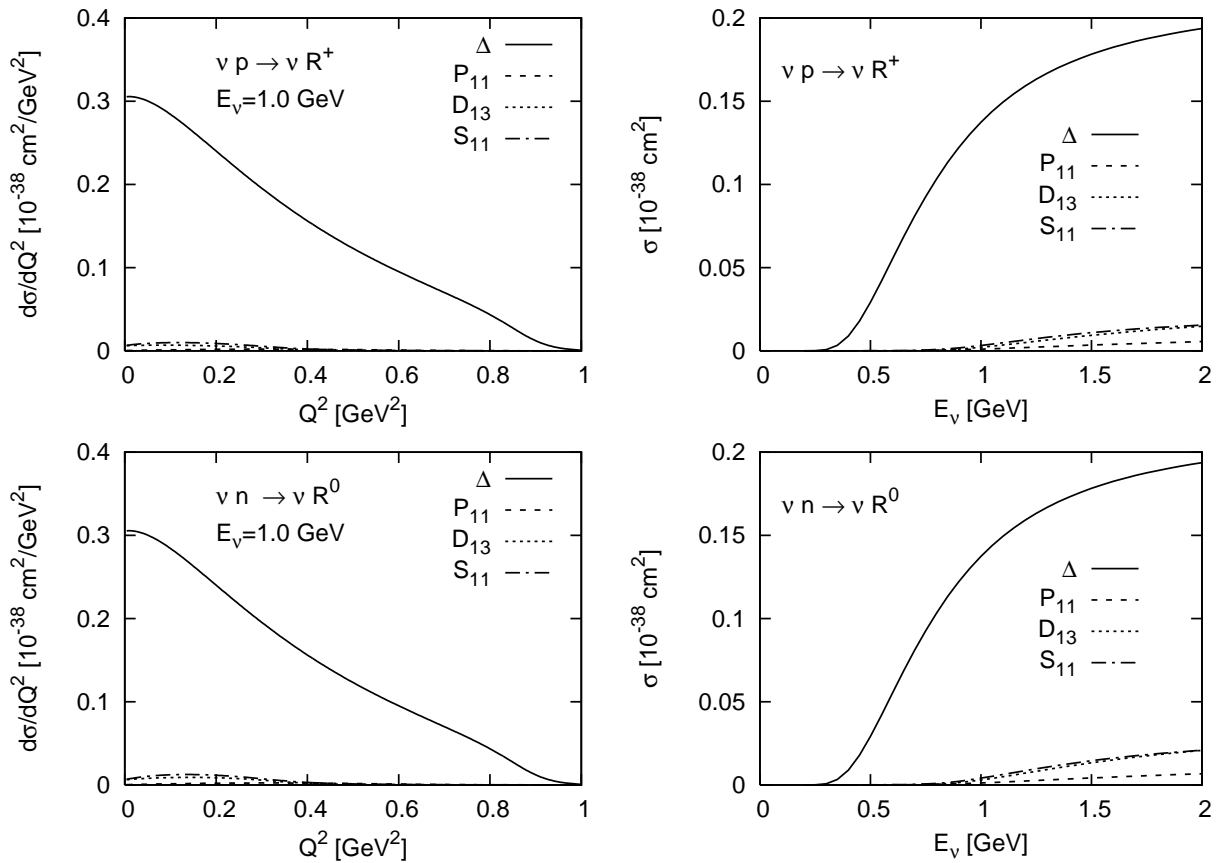}
   \caption{Differential and integrated cross section for NC resonance excitation on protons (top) and neutrons (bottom). The cross section for $\Delta$ excitation (solid lines) clearly dominates. The lines denoting the cross section for $D_{13}$ overlap with the ones for the $S_{11}$.  \label{fig:nuRES}}
\end{figure*}

The cross sections for NC resonance excitation on protons and neutrons are shown in \reffig{fig:nuRES}. In both cases, the $\Delta$ yield is considerably larger than the other ones. 

\begin{figure*}
\includegraphics[width=16cm]{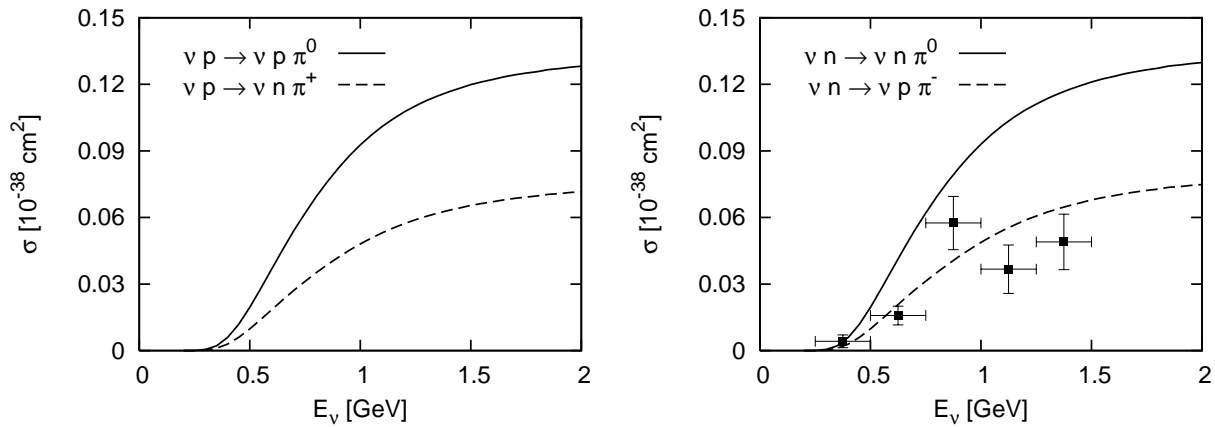}
   \caption{Integrated cross section for NC one-pion production via resonance excitation and subsequent decay. The left panel shows the pion production on protons and the right one on neutrons. Data are shown for the reaction $\nu n \to \nu p \pi^-$ (taken from \refcite{Derrick:1980nr}). \label{fig:nucl_pion_prod}}
\end{figure*}

In the vacuum, we model pion production via resonance excitation and their subsequent decay (we emphasize that in the nuclear medium resonances can undergo "pionless decay" (cf.~\refch{ch:inmedFSI})). 
When a neutrino scatters with a nucleon, either an isospin $3/2$ resonance ($\Delta$) or an isospin $1/2$ resonance ($P_{11}$, $D_{13}$ and $S_{11}$) is produced. They decay, among other channels, with a certain branching ratio into $\pi N$ pairs. By including the appropriate Clebsch-Gordan coefficients we obtain for the cross sections of the four possible one-pion production channels
\bea
\sigma (\nu p \to \nu n \pi^+) &=& \frac13 \sigma_{\Delta^+} + \frac23 \left(b_1 \sigma_{P_{11}^+} +  b_2 \sigma_{D_{13}^+} +  b_3 \sigma_{S_{11}^+}  \right), \\
\sigma (\nu p \to \nu p \pi^0) &=& \frac23 \sigma_{\Delta^+} + \frac13 \left(b_1 \sigma_{P_{11}^+} +  b_2 \sigma_{D_{13}^+} +  b_3 \sigma_{S_{11}^+}  \right), \\
\sigma (\nu n \to \nu p \pi^-) &=& \frac13 \sigma_{\Delta^0} + \frac23 \left(b_1 \sigma_{P_{11}^0} +  b_2 \sigma_{D_{13}^0} +  b_3 \sigma_{S_{11}^0}  \right), \\
\sigma (\nu n \to \nu n \pi^0) &=& \frac23 \sigma_{\Delta^0} + \frac13 \left(b_1 \sigma_{P_{11}^0} +  b_2 \sigma_{D_{13}^0} +  b_3 \sigma_{S_{11}^0}  \right), 
\eea
where $\sigma_{R^+}$ ($\sigma_{R^0}$) are the cross sections for resonance excitation on protons (neutrons) as shown in \reffig{fig:nuRES}. For consistency, the branching ratios $b_i$ are also taken from the analysis of Manley and Saleski \cite{manley} which gives $b_1=0.69$, $b_2=0.59$ and $b_3=0.51$.  
In \reffig{fig:nucl_pion_prod}, we plot the results for the integrated pion production cross section on protons (left) and neutrons (right). 
Data on NC pion production is extremely sparse. A measurement was performed on D$_2$ at the ANL bubble chamber for the $\nu n \to \nu p \pi^-$ channel~\cite{Derrick:1980nr} --- these data are also shown in \reffig{fig:nucl_pion_prod}. The remaining channels have only been measured at the Gargamelle bubble chamber~\cite{Krenz:1977sw} (see also the reanalysis by Hawker~\cite{hawker}) on a propane-freon mixture and not on "elementary targets". Thus, we do not show them here. 
However, we have compared our results to the calculations of Paschos \refetal{Paschos:2000be} (cf.~their Figs.~4~-~7) and have found reasonable agreement for the integrated vacuum one-pion production cross sections.

\section{Neutrino-nucleus reactions}

\subsection{In-medium modifications and final-state interactions \label{ch:inmedFSI}}

Nuclear effects play the central role in this study. A detailed description of our nuclear model is given 
  in \refcite{Leitner:2006ww} where we applied it to CC neutrino-nucleus scattering. Here we only outline the 
main features. 

The cross section for neutrino scattering on free nucleons, given in \refeq{eq:crosssection}, has to be modified when the nucleon is bound in the nucleus. First, we consider Fermi motion of the initial nucleons and Pauli blocking of the final ones in a local density approximation, in which the local Fermi momenta of the nucleons are 
given by 
\be
p_F(\myvec{r})=\left( \frac{3}{2} \pi^2 \rho(\myvec{r}) \right)^{1/3}.
\ee
For the density distribution of heavier nuclei we use a Woods-Saxon form with parameters extracted from Hartree-Fock calculations. 
For $^{12}$C, we take a harmonic oscillator density as given in \refcite{Nieves:1993ev}. 

The nucleons are bound in a density and momentum dependent scalar potential $U(\myvec{r},\myvec{p})$,
whose parametrization has been obtained by fits to the saturation density of nuclear matter, and also to the momentum dependence of the nucleon optical potential as measured in $pA$ collisions~\cite{Effenberger:1999ay}. 
Since the $\Delta$ is less bound in the nucleus than the nucleons --- at normal nuclear density one finds $U_{\Delta} \approx -30 \myunit{MeV}$ versus $U_N \approx -45 \myunit{MeV}$~\cite{Ericson:1988gk} ---, we set the $\Delta$ potential to $2/3$ of the nucleon potential. For the other resonances the same potential as for the nucleons is adopted.
A particle bound in the nucleus acquires an effective mass, defined as
\be
M_{eff}=M_{N,R} + U_{N,R}(\myvec{r},\myvec{p}),  
\ee
where $M_{N,R}$ is the corresponding "free" mass.
As consequence of the density and momentum dependence of the scalar potential the effective masses of initial and final particles in a scattering process can be different even if their rest masses are equal. We account for this fact by replacing $M$ and $M'$ (note that $M_N$ remains unchanged) in the cross section formula (\refeq{eq:crosssection}) and in the hadronic tensors (\refeq{eq:hadrtensorQE} and \refeq{eq:hadrtensorDELTA}) by their respective effective masses (cf. \refcite{Leitner:2006ww} for details).

The final-state interactions (FSI) of the produced particles are implemented by means of the coupled-channel semi-classical Giessen Boltzmann-Uehling-Uhlenbeck (GiBUU) transport model~\cite{gibuu}. Originally developed to describe heavy-ion collisions~\cite{Teis:1996kx, Hombach:1998wr, Wagner:2004ee}, it has been extended to describe the interactions of pions, photons and electrons with nuclei~\cite{Weidmann:1997vj, Effenberger:1999ay, Lehr:1999zr, Falter:2004uc, Buss:2006vh, Buss:2006yk}. Recently, we further extended the GiBUU model to describe neutrino scattering off nuclei~\cite{Leitner:2006ww}. This extension does not require the introduction of any new nuclear parameter.

In our model, the space time evolution of a many-particle system under the influence of a mean-field potential and a collision term is described by a BUU equation for each particle species. A collision term accounts for changes (gain and loss) in the phase space density due to elastic and inelastic collisions between the particles and also due to particle decays into other hadrons whenever it is allowed by Pauli blocking. In between the collisions, all particles (also resonances) are propagated in their mean-field potential according to their BUU equation (cf.~\refscite{Effenberger:1999ay,Falter:2004uc}). More details on the cross sections, their in-medium modification and the included particles are given in the aforementioned references. 

Inside the nuclear medium, the width of the resonances is modified. Since their decay into $\pi N$ pairs might be Pauli blocked, their width is lowered compared to the vacuum width. On the contrary, we have an increase of the width due to collisions in the medium. Elastic and inelastic scattering of the resonances with the nucleons contribute to this collisional broadening. Therefore, the total in-medium width is given by a sum of the Pauli modified decay width~$\tilde{\Gamma}$ and the collisional width~$\Gamma_{coll}$. Despite the decrease caused by Pauli blocking, the dominant effect is a broadening of the resonances in the medium.

The most relevant states for neutrino-induced reactions at intermediate energies are the nucleon, the $\Delta$~resonance, the pion and their interactions. For the $NN$ cross section and its angular dependence we use a fit to data from \refcite{Cugnon:1996kh}. For the pion cross sections we use a resonance model with the background fitted to data as shown in detail in \refcite{Buss:2006vh}. 
The $\Delta$~resonance is propagated off-shell in our model and decays isotropically. Its decay into a pion nucleon pair is Pauli blocked if the momentum of the nucleon is below the Fermi momentum. We allow not only for the decay of the $\Delta$, but also for FSI of the $\Delta$ in the nuclear medium. Possible absorption processes include $\Delta N \to N N$, $\Delta N N \to N N N$ and $\Delta N \to \pi N N$. All those processes are implemented by an absorption probability depending on the in-medium collisional width for which we use the results of Oset and Salcedo~\cite{Oset:1987re}. We emphasize that the whole $\pi N \Delta$ dynamics has been tested extensively and also compared to data in $A \, A \to \pi \, X$~\cite{Teis:1996kx}, $\pi \, A \to X$~\cite{Engel:1993jh, Buss:2006vh} and $\gamma \, A \to \pi \, X$~\cite{Lehr:1999zr} and recently in pionic double charge exchange reactions \cite{Buss:2006yk}.

In conclusion, FSI lead to absorption, charge exchange, a redistribution of energy and momentum as well as to the production of new particles. In our coupled-channel treatment of the FSI --- in which the BUU equations are coupled through the collision term and, with less strength, also through the potentials --- our model differs from standard Glauber approaches because the collision term allows not only for absorption but also for side-feeding and rescattering. 
In addition, our model is applicable in many different nuclear reactions using the same physics input. In particular, an important prerequisite for any model aiming at the description of the interaction of neutrinos with nuclei is a good description of electron- or photon-induced reactions. There, extensive tests against existing data are possible and have been successfully performed~\cite{Falter:2003uy, Alvarez-Ruso:2004ji}. 

\subsection{Results and discussion}
We now present our results for NC pion production and nucleon knockout for neutrino energies up to 2~GeV for nuclei commonly used in neutrino experiments.  

\subsubsection{Pion production}
\begin{figure*}
  \begin{minipage}{.78\textwidth}
      \begin{center}
        \includegraphics[height=5.5cm]{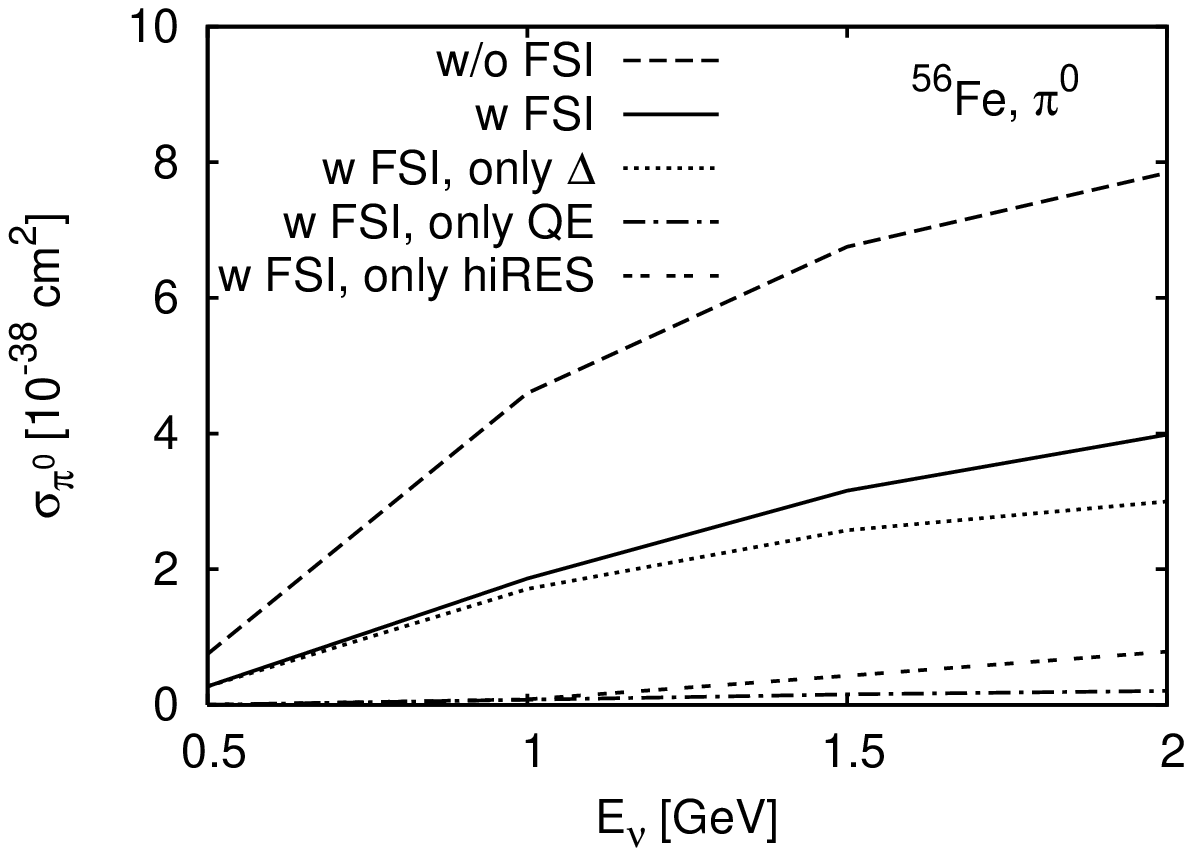}
       \end{center}
  \end{minipage}
 \\ \hfill \\ \hfill \\
  \begin{minipage}{.48\textwidth}
      \begin{center}
        \includegraphics[height=5.5cm]{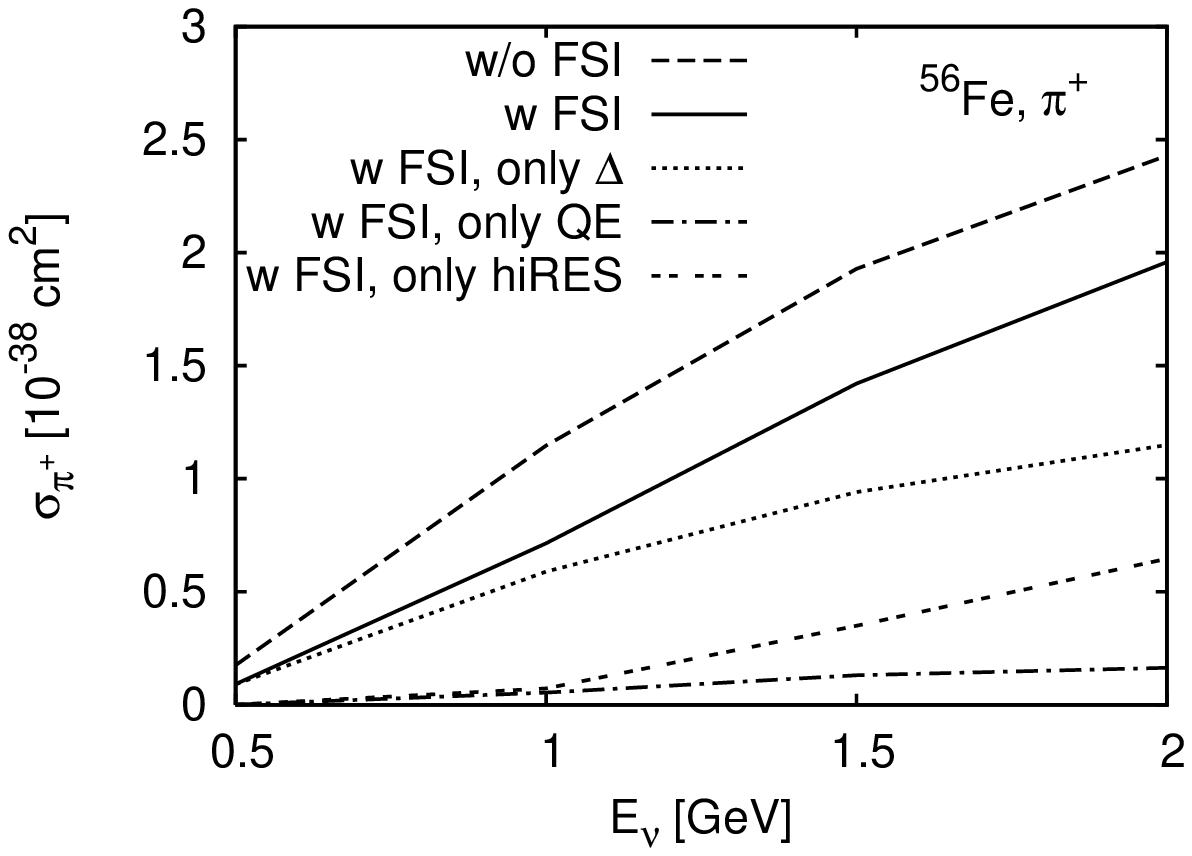}
      \end{center}
  \end{minipage}
  \begin{minipage}{.48\textwidth}
      \begin{center}
        \includegraphics[height=5.5cm]{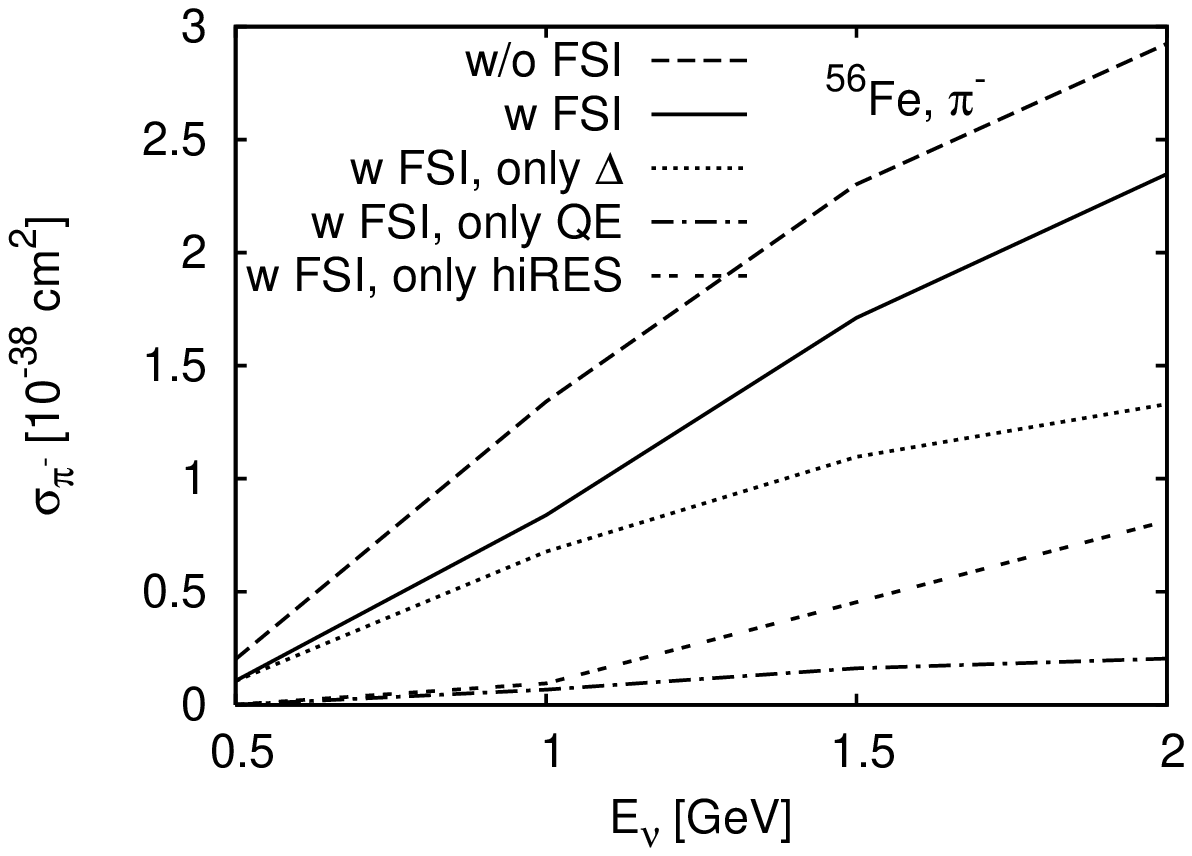}
      \end{center}
  \end{minipage}
\caption{Integrated cross section for $\pi^0$ (top), $\pi^+$ (bottom left) and $\pi^-$ (bottom right) production on $^{56}\text{Fe}$ as a function of $E_{\nu}$. The dashed lines show the results without FSI (only the decay of resonances is possible); the results denoted by the solid lines include FSI. Also indicated is the origin of the pions (QE, $\Delta$ excitation or higher resonances (hiRES)). \label{fig:pion_sigmatot}}  
\end{figure*}

We begin our discussion on pion production with the total production cross sections for $\pi^+$, $\pi^0$ and $\pi^-$ on $^{56}$Fe which are shown in \reffig{fig:pion_sigmatot}. 
The dashed line denotes the pions stemming from the decay of the initially produced resonances --- no further FSI are taken into account here. The cross section for $\pi^0$~production is significantly higher than those of the charged channels. This is a direct consequence of the isospin structure of the resonance decay. "Switching on" FSI allows the resonances to interact in different ways besides a simple decay. Also the produced pions can interact further or undergo absorption. 
These FSI lead to a strong reduction of the total yield in the $\pi^0$~channel (compare solid and dashed line in the upper panel). The reduction is much smaller in the $\pi^+$ and $\pi^-$~channel because the $\pi^0$ undergo charge exchange scattering and thus contribute to the charged channels (side-feeding). The effect in the opposite direction is less important due to the smaller elementary $\pi^+$ and $\pi^-$ production cross section. 
Thus, side-feeding shifts strength always from the dominant into the less dominant channel. In the context of neutrino reactions, this effect was first described by Adler \refetal{Adler:1974qu}. It is also observed in charged currents from the $\pi^+$ into the $\pi^0$ channel \cite{Paschos:2003ej, Paschos:2004qh, Leitner:2006ww}. 

Pions can also emerge from the initial QE neutrino-nucleon reaction when the produced nucleon rescatters producing a $\Delta$ or directly a pion. However, as can be seen from \reffig{fig:pion_sigmatot}, this process is not very sizable because it is relevant only at high $Q^2$ (dash-dotted lines). 

\begin{figure*}
    \includegraphics[width=17cm]{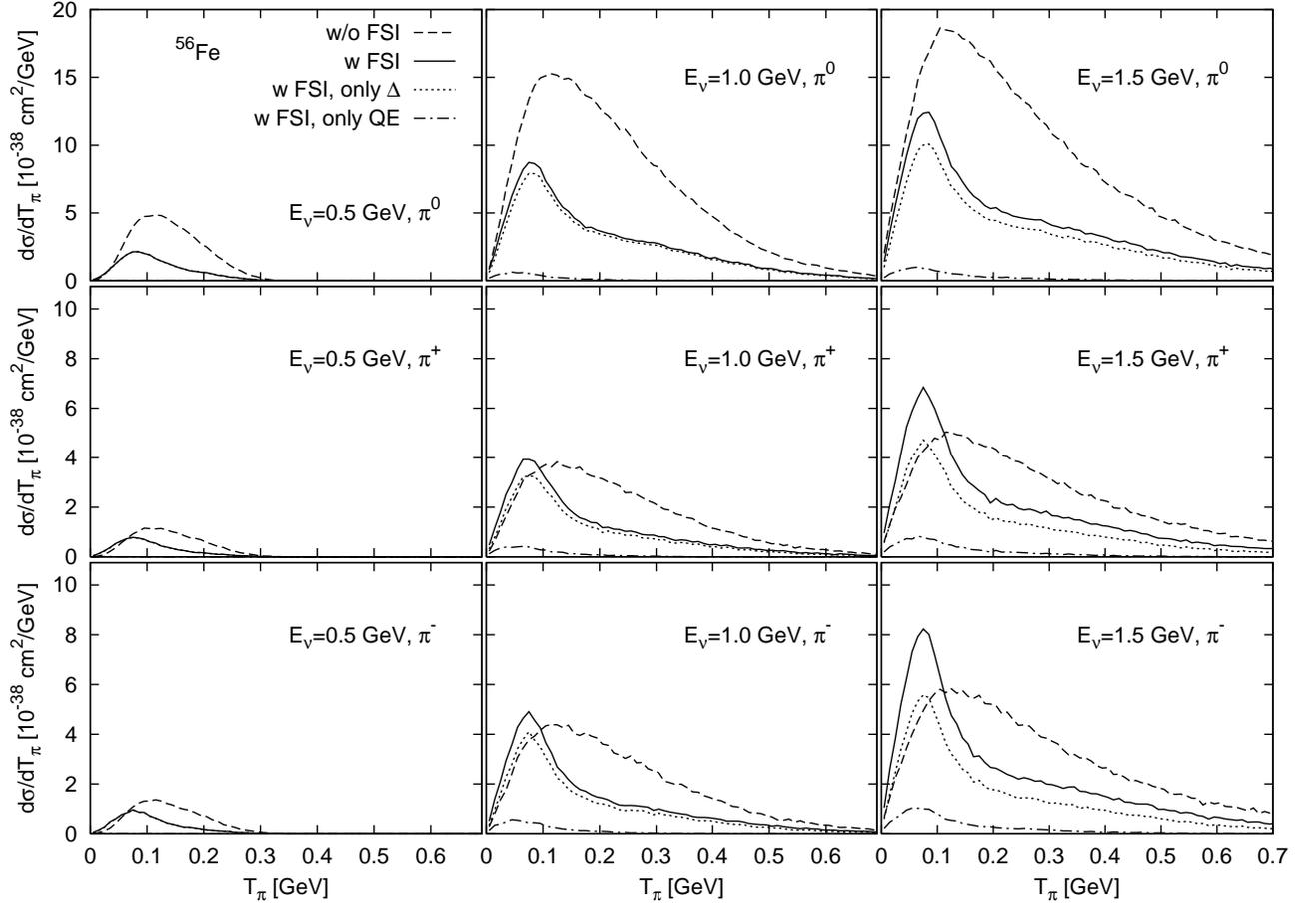}
   \caption{Kinetic energy differential cross section for $\pi$ production on $^{56}\text{Fe}$ versus the pion kinetic energy $T_{\pi}$ at different values of $E_{\nu}$. The dashed lines denote the calculation without FSI where only the decay of resonances is included; the solid lines denote the one with FSI. Also indicated is whether the pion comes from initial QE or $\Delta$ excitation (dash-dotted or dotted lines).\label{fig:pion_diff}}
\end{figure*}

A more detailed understanding of pion production is obtained by studying the pion kinetic energy distributions. In \reffig{fig:pion_diff} we show the kinetic energy spectra for $\pi^0$, $\pi^+$ and $\pi^-$ production for three neutrino energies. The dashed lines denote again the result without final-state interactions and the solid lines the result of the full calculation. The contributions from initial $\Delta$ excitation (dotted lines) and from initial QE events (dash-dotted lines) are also shown. Pion production through FSI of initial QE processes contributes mostly to the low energy region of the pion spectra due to the redistribution of the energy in the collisions. While the overall shape of the dashed lines (without FSI) is dictated by the predominant $p$-wave production mechanism through the $\Delta$ resonance, the shape of the solid lines (full calculation) is influenced by the energy dependence of the pion absorption and rescattering. The main absorption mechanism for pions above $T_{\pi}\approx 0.1$~GeV is $\pi N \to \Delta$ followed by $\Delta N \to N N$ which leads to a considerable reduction of the cross section. Elastic scattering $\pi N \to \pi N$ redistributes the kinetic energies and thus also shifts the spectrum to lower energies. In the case of the smaller $\pi^+$ and $\pi^-$ channels the already discussed side-feeding enhances the peak in the middle and bottom panels of \reffig{fig:pion_diff} over the value obtained without FSI. Note that these spectra are very similar in shape to those measured in $(\gamma,\pi^0)$ reactions on nuclei (cf.~Figs.~13 and 14 in \refcite{Krusche:2004uw}).

\begin{figure*}
      \begin{minipage}{.48\textwidth}
        \begin{center}
          \includegraphics[height=5.5cm]{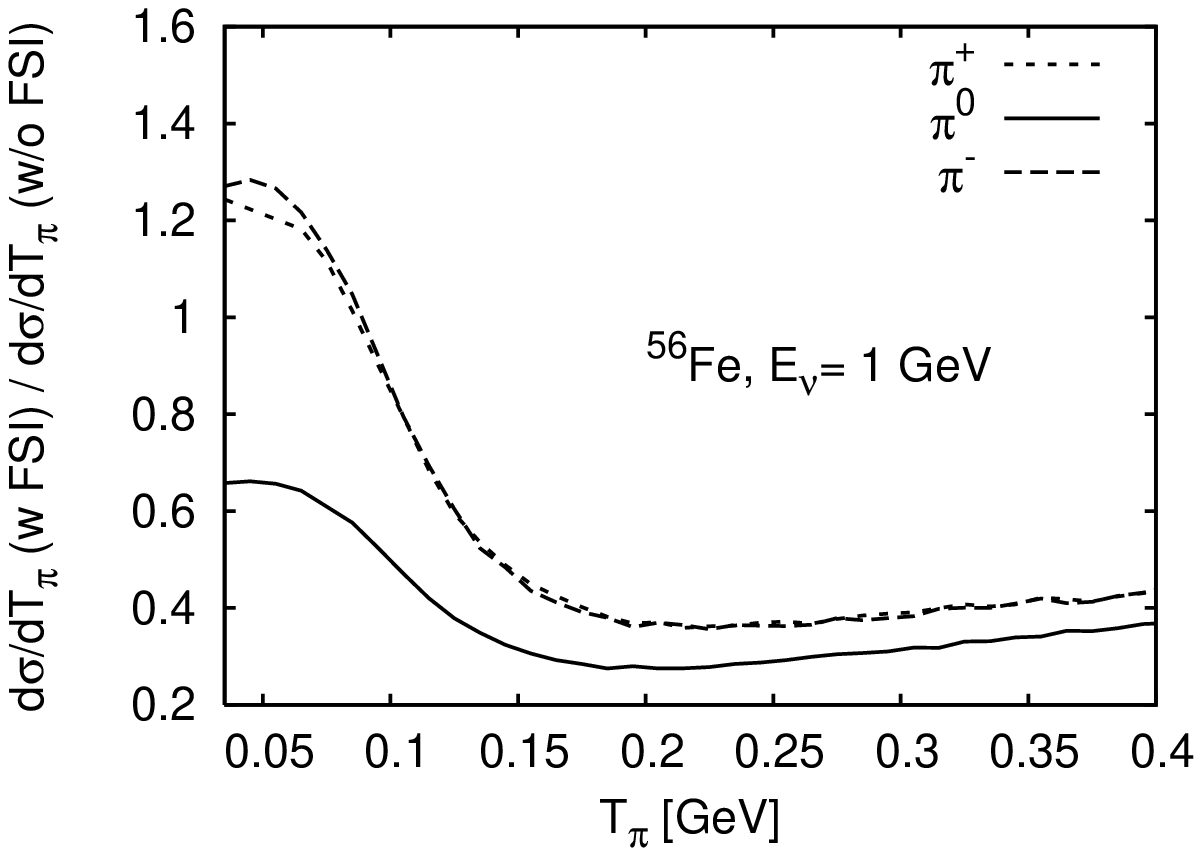}
           \end{center}
    \end{minipage}
      \begin{minipage}{.48\textwidth}
        \begin{center}
          \includegraphics[height=5.5cm]{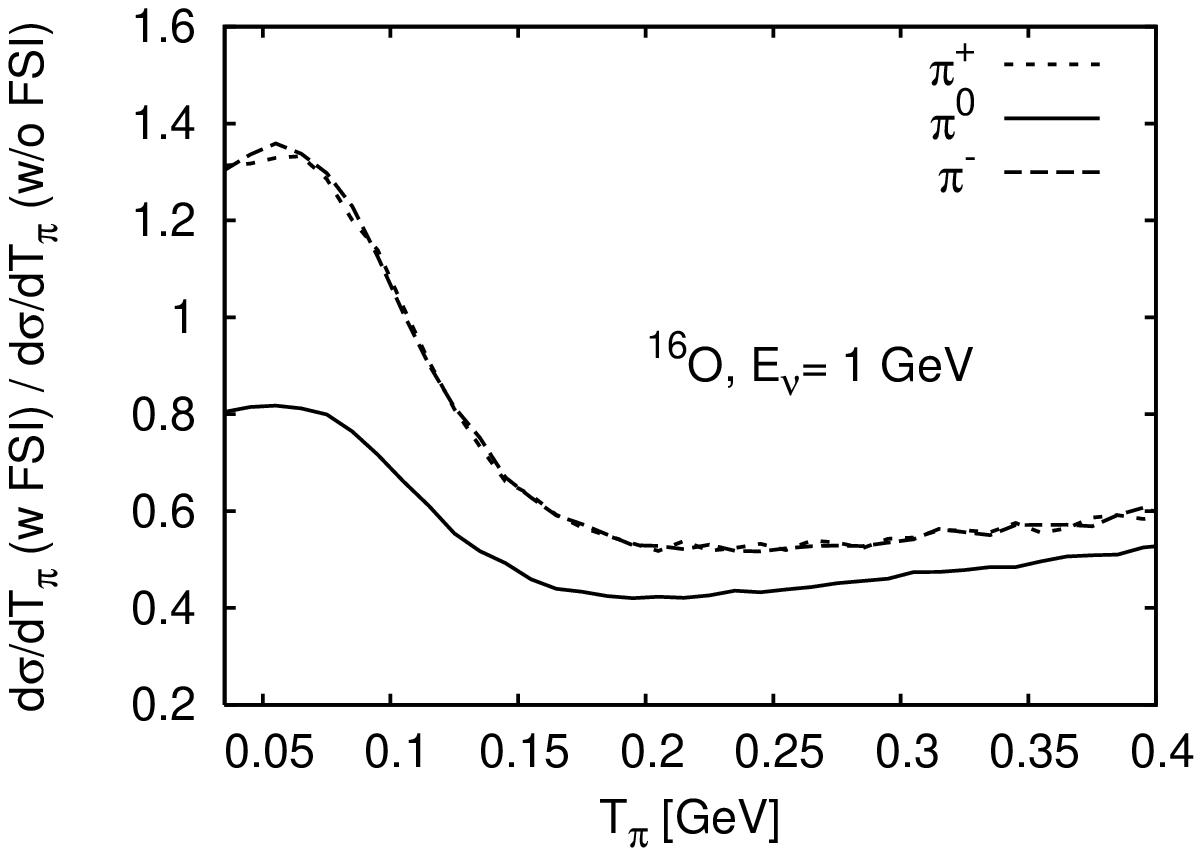}
        \end{center}
    \end{minipage}
    \caption{Ratio of the differential cross sections (the cross section with FSI divided by the one without FSI) for pion production on $^{56}\text{Fe}$ (left) and $^{16}\text{O}$ (right) versus the pion kinetic energy at $E_{\nu}= 1 \myunit{GeV}$. The initial QE scattering process has been "switched off" here. 
    \label{fig:ratio}}
    \end{figure*}

The impact of FSI is even more visible in the ratios obtained by dividing the differential cross section with FSI by the one without FSI. These are plotted in \reffig{fig:ratio} for $^{56}$Fe and $^{16}$O at $E_{\nu}=1$~GeV versus the pion kinetic energy (short-dashed for $\pi^+$, solid for $\pi^0$ and dashed lines for $\pi^-$). As expected, the absorption is bigger in the heavier nucleus ($^{56}\text{Fe}$) than in the lighter one ($^{16}\text{O}$) --- it scales with~$A^{2/3}$. 
For pions with kinetic energy $\gtrsim 0.1$~GeV we find strong effects of FSI.
This is the region where pion absorption and rescattering are most prominent due to the excitation of the $\Delta$ resonance around its peak position. At lower energies ($T_{\pi} \approx 0.07$~GeV) we find a peak because pions of higher energy in average lose energy via rescattering. At still lower pion energies, the multi-nucleon pion absorption mechanism takes over, leading to a small dip. We stress that a similar pattern has been experimentally observed by Krusche \refetal{Krusche:2004uw} in pion photoproduction (cf.~Fig.~16 in \refcite{Krusche:2004uw}). This particular dependence of the ratio reflects well-known features of the $\pi N \Delta$ dynamics in nuclei. 

It would be interesting to compare these results with earlier calculations of neutrino-induced pion production on nuclei by Paschos \refetal{Paschos:2000be}. However, these authors have very recently found an error in their elementary pion-production spectra~\cite{paschosprivcomm} which affects their Figs. 8-14 for the pion energy distribution in \refcite{Paschos:2000be}. In view of this we abstain from a comparison until the corrected results are published.

\subsubsection{Nucleon knockout}

\begin{figure*}
  \begin{minipage}{.48\textwidth}
      \begin{center}
        \includegraphics[height=5.5cm]{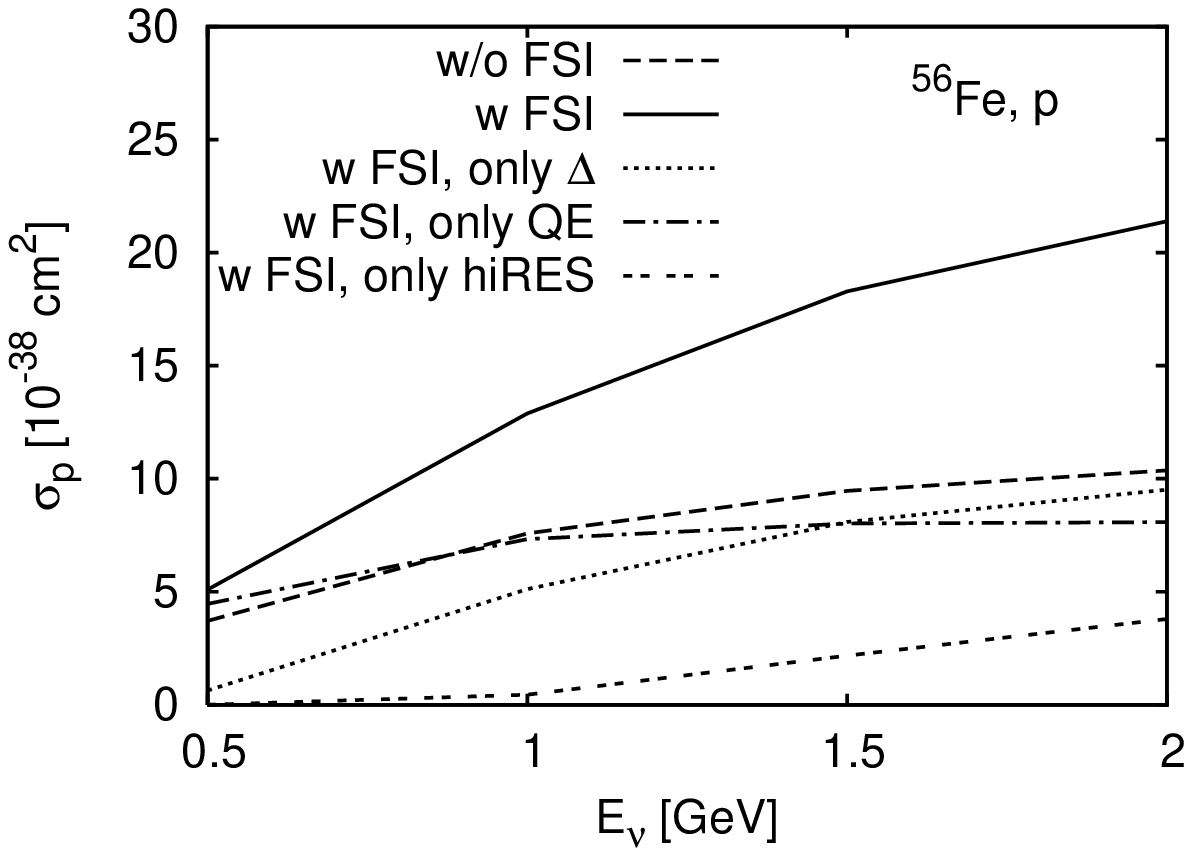}
      \end{center}
  \end{minipage}
  \begin{minipage}{.48\textwidth}
      \begin{center}
        \includegraphics[height=5.5cm]{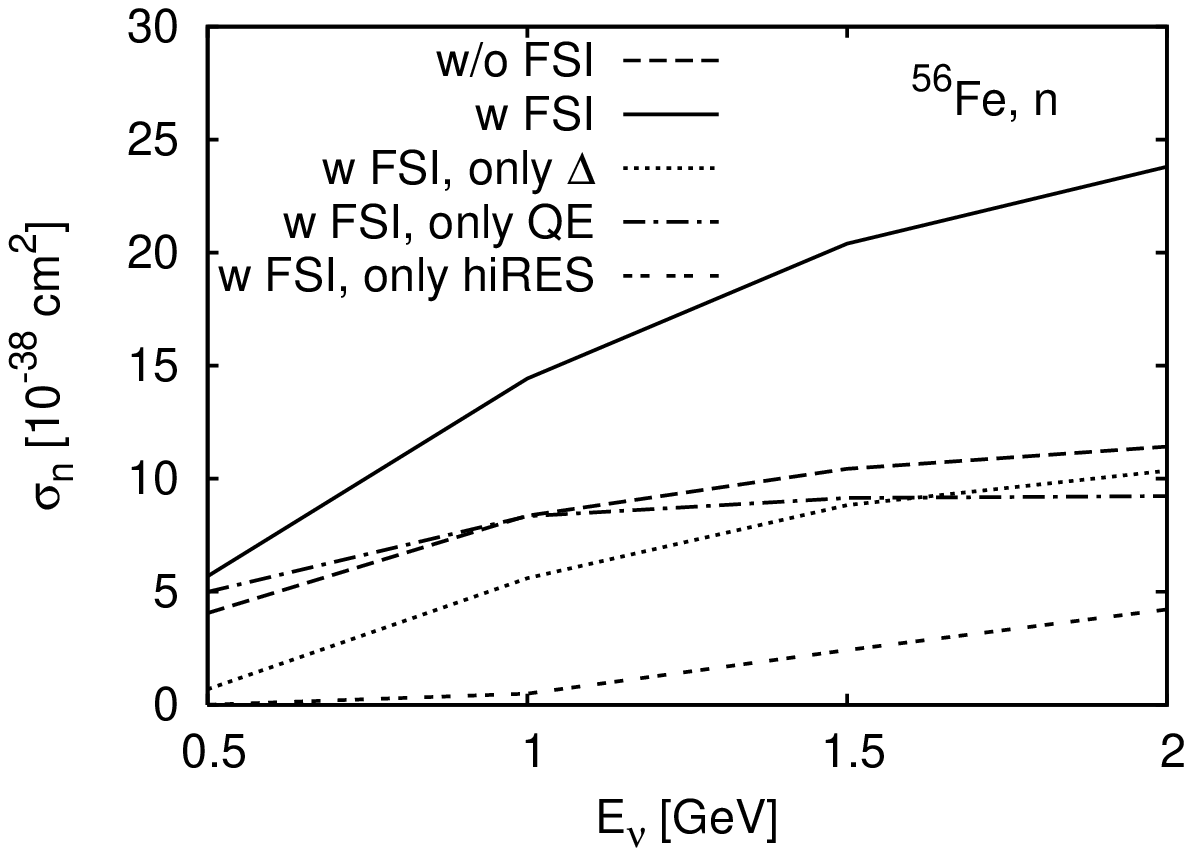}
      \end{center}
  \end{minipage}
\caption{Integrated cross section for proton (left) and neutron (right) knockout on $^{56}\text{Fe}$ versus $E_{\nu}$. The dashed lines show the results without FSI (only the decay of resonances is possible); the results denoted by the solid lines include FSI. Also indicated is the origin of the nucleons (QE, $\Delta$ excitation or higher resonances (hiRES)). \label{fig:nucl_sigmatot}}  
\end{figure*}

We now continue our discussion on exclusive channels with nucleon knockout. We consider all nucleons leaving the nucleus in the $\nu A$ reaction which --- in case FSI are included --- is not necessarily a single-nucleon knockout.
In \reffig{fig:nucl_sigmatot} we show the integrated cross sections for proton and neutron knockout on $^{56}$Fe. The solid lines, showing the results with FSI included, lie in both cases clearly above the ones without FSI (dashed lines). This enhancement is entirely caused by secondary interactions.
Furthermore, the initial process leading to a knocked out nucleon is indicated. In the $\nu N$ collision either QE scattering (dash-dotted), $\Delta$ (dotted) or $N^*$ excitation (double-dashed lines) is possible. 
While the pion cross section was dominated by the initial $\Delta$ excitation, here initial QE and initial $\Delta$ excitation contribute to the total cross section above $E_{\nu} \approx 1$~GeV with nearly equal amounts. This reflects the energy dependence of the $\nu N$ cross sections (cf.~\reffig{fig:nuQE} and \ref{fig:nuRES}).
Only up to neutrino energies of $\approx 0.5$~GeV the resonance contributions to nucleon knockout is negligible. We note that our method of using the $\Delta$ selfenergies of Oset and Salcedo~\cite{Oset:1987re} (cf.~\refch{ch:inmedFSI}) even underestimates the number of knocked out nucleons because we do not treat the process $\Delta N \to NN$ explicitely. Our $\Delta$ resonance contribution of knocked out nucleons stems solely from the decay $\Delta \to \pi N$.

\begin{figure*}
    \includegraphics[width=15.5cm]{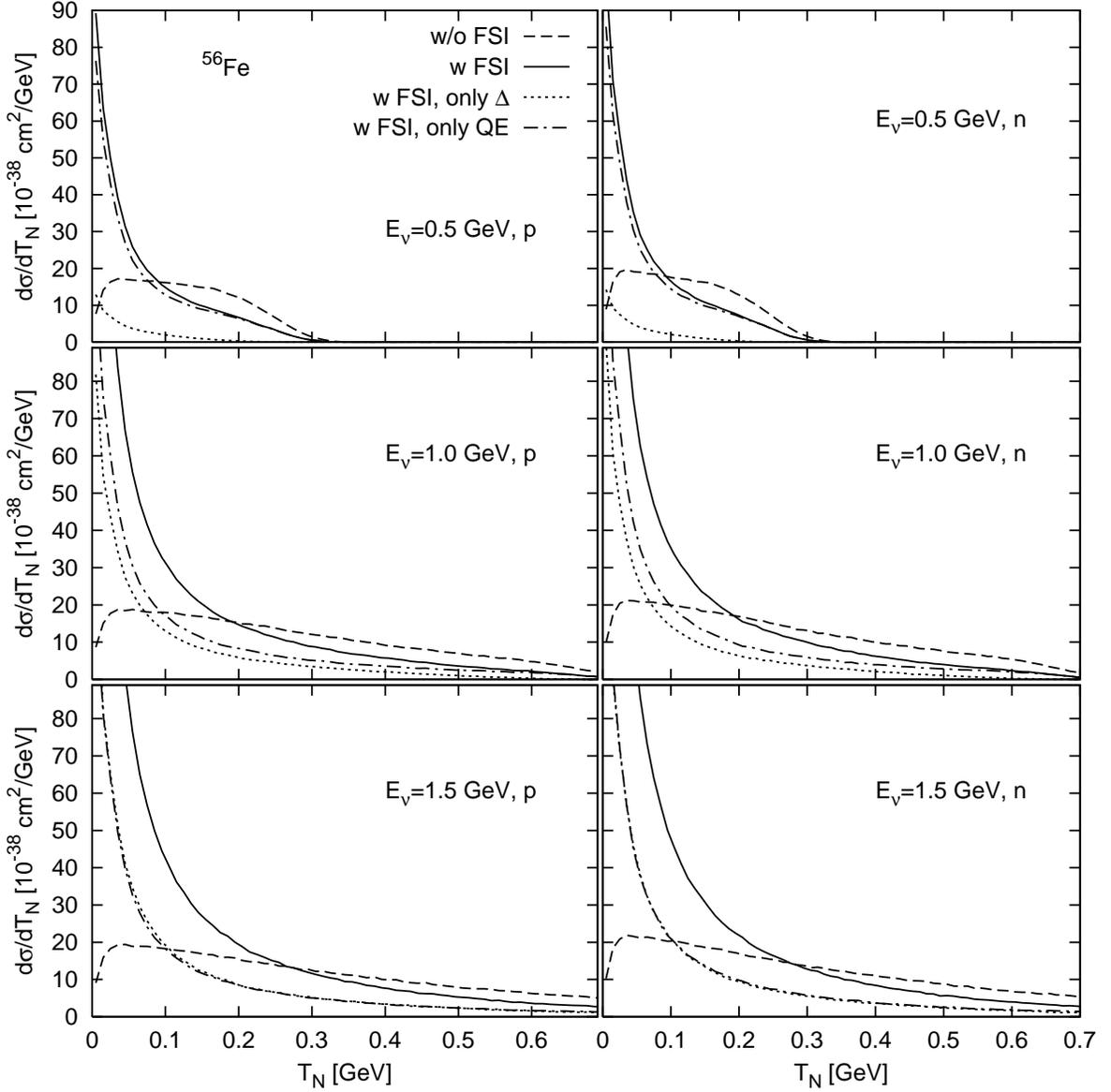}
   \caption{Kinetic energy differential cross section for proton (left) and neutron (right) knockout on $^{56}\text{Fe}$ versus the nucleon kinetic energy $T_{N}$ at different values of $E_{\nu}$. The dashed lines denote the calculation without FSI where only the decay of resonances is included; the results denoted by the solid lines are with FSI. Also indicated is whether the pion comes from initial QE or $\Delta$ excitation (dash-dotted or dotted line). \label{fig:nucleon_diff}}
\end{figure*}

In \reffig{fig:nucleon_diff} we plot the kinetic energy differential cross section for proton and neutron knockout versus the kinetic energy on $^{56}$Fe for different values of $E_{\nu}$. The inclusion of FSI strongly modifies the shape of the distribution (compare the dashed and the solid lines). This effect is due to the rescattering of high energy nucleons in the medium which reduces the flux at higher energies while, simultaneously, a large number of secondary nucleons at lower energies is emitted. At $E_{\nu}=0.5$~GeV, nucleon knockout is clearly dominated by QE processes, however, the plots show how the $\Delta$ becomes progressively more important as $E_{\nu}$ increases.

\begin{figure*}
    \includegraphics[width=16cm]{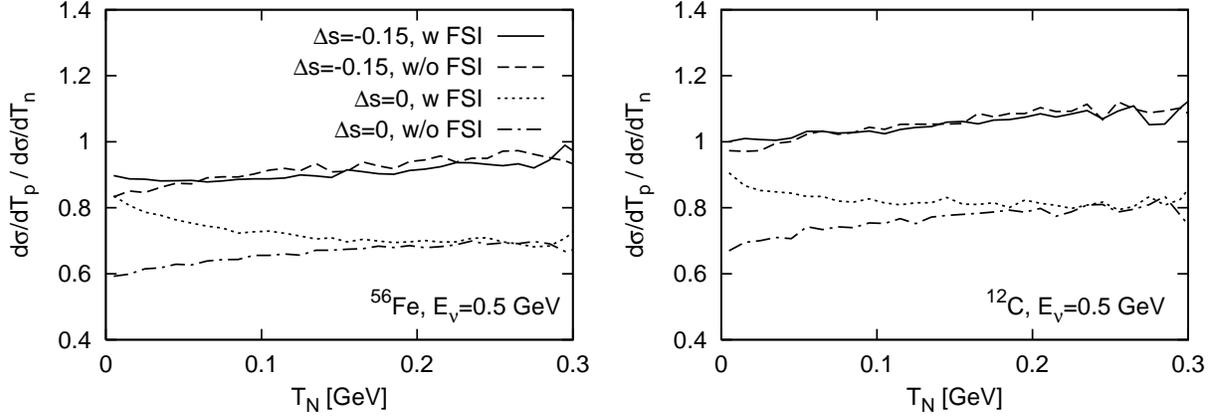}
   \caption{Ratio of proton to neutron kinetic energy differential cross section on $^{56}\text{Fe}$ (left) and $^{12}\text{C}$ (right) versus the nucleon kinetic energy $T_{N}$ at $E_{\nu}=0.5$~GeV. The solid and the long-dashed lines show the results with and without FSI for $\Delta s = -0.15$, the dotted and the dash-dotted lines the results for $\Delta s = 0$. \label{fig:influence_deltas_ratio}}
\end{figure*}

In contrast to the CC reaction studied in \refcite{Leitner:2006ww}, where neutrons were only emitted by secondary collisions, in the NC case, both the neutron and proton kinetic energy distributions are equally affected by FSI since their total yields without FSI are comparable. 
This can be seen also in \reffig{fig:influence_deltas_ratio} where we plot the ratio of the proton to neutron kinetic energy differential cross section for $^{56}\text{Fe}$ and $^{12}\text{C}$ at $E_{\nu}=0.5$~GeV. The calculations with and without FSI (solid and dashed lines) agree approximately if $\Delta s=-0.15$. This shows that the effect of the final-state interaction cancels out regardless of the nucleus. This is different in the CC case (cf.~\refcite{Leitner:2006ww}) where we found strong effects of side-feeding from the dominant $p$ channel into the suppressed $n$ channel. In fact, side-feeding is only important when the initial proton and neutron yields are different.
Indeed, when we "switch off" the strange axial form factor (i.~e.~take $\Delta s=0$) then the elementary proton and neutron yields are different as shown in \reffig{fig:nuQE}; while the neutron cross section is enhanced, the proton cross section is reduced. Therefore, we expect side-feeding from the neutron to the proton channel which changes the $p/n$ ratio as observed in \reffig{fig:influence_deltas_ratio} (dash-dotted versus dotted lines). Since nucleons lose energy in (charge exchange) scattering, the effect is more pronounced at low kinetic energies.

\begin{figure*}
  \begin{minipage}{.48\textwidth}
      \begin{center}
        \includegraphics[height=5.5cm]{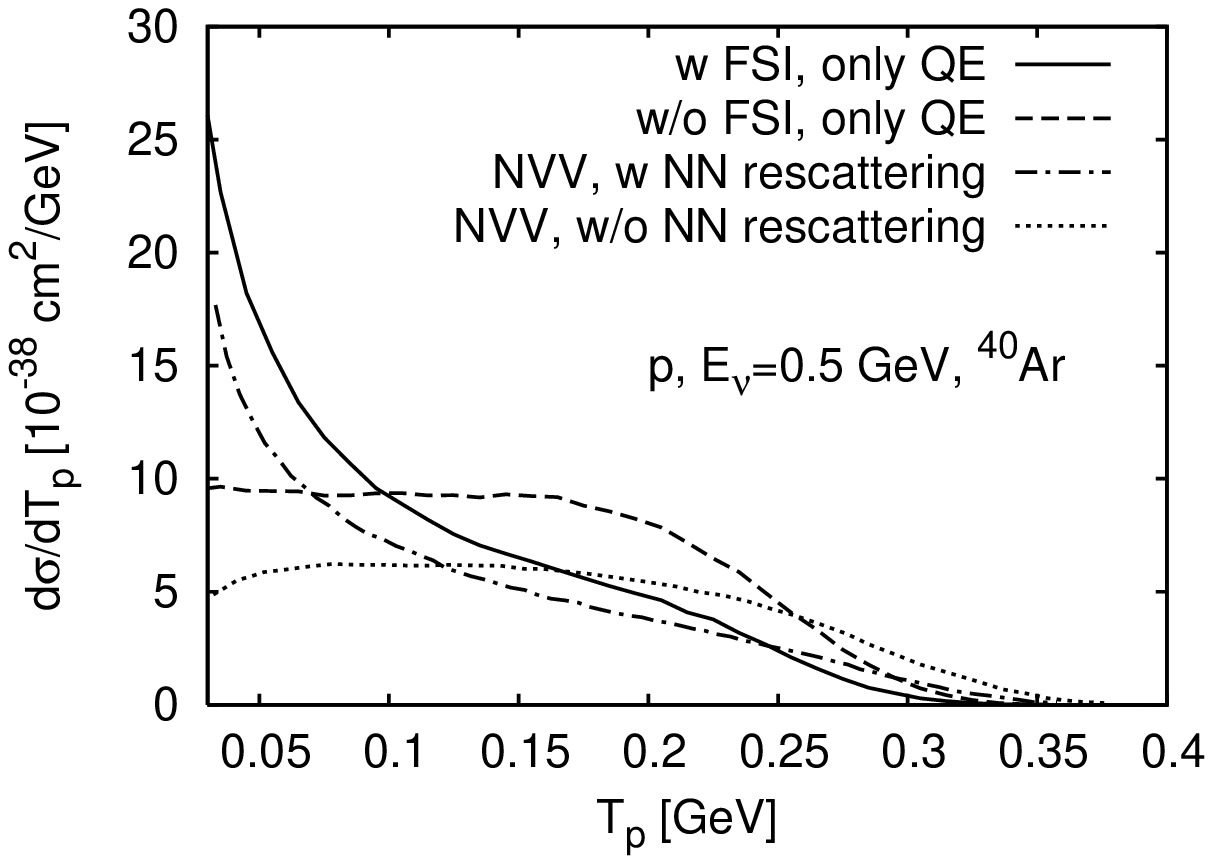}
      \end{center}
  \end{minipage}
  \begin{minipage}{.48\textwidth}
      \begin{center}
        \includegraphics[height=5.5cm]{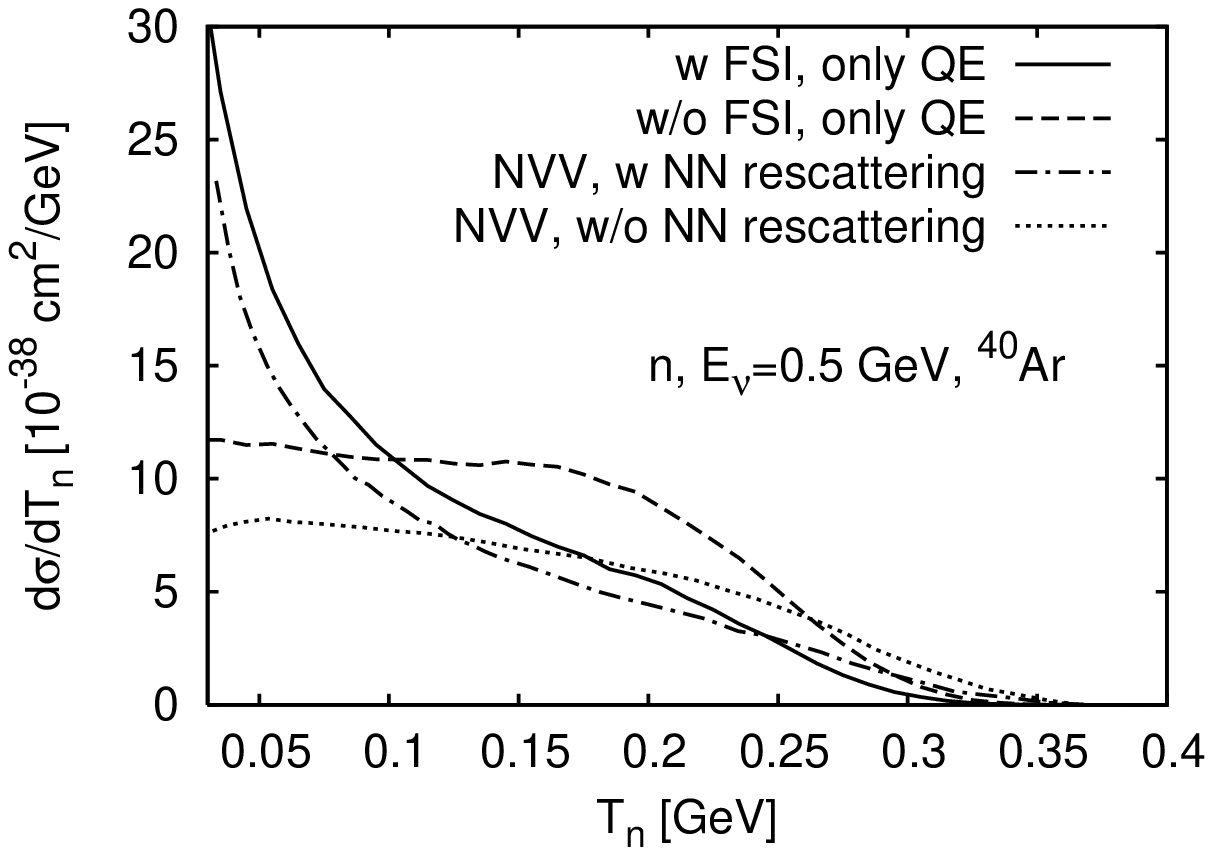}
      \end{center}
  \end{minipage}
   \caption{Differential cross section for neutrino induced proton and neutron knockout on $^{40}$Ar at $E_{\nu}=0.5 \myunit{GeV}$. The initial neutrino-nucleon resonance production processes have been "switched off" in this case. The dashed lines denote our result without FSI and the solid lines the one with FSI. The dotted (dash-dotted) lines denoted with "NVV" show the results of Nieves \refetal{Nieves:2005rq} without (with) $NN$ rescattering. \label{fig:nieves_comp}}
\end{figure*}

Finally, we compare our results on nucleon knockout with other calculations. In \reffig{fig:nieves_comp} we show our results for the kinetic energy differential cross section together with those of Nieves \refetal{Nieves:2005rq} (denoted with "NVV"). Since they do not include any resonances, we have "switched off" the initial resonance excitation in our calculation, so only nucleon knockout induced by initial QE events is considered. The discrepancy of our result without FSI to the result without $NN$ rescattering of \refcite{Nieves:2005rq} (dashed versus dotted line in \reffig{fig:nieves_comp}) could be attributed to  --- in addition to differences in the momentum distribution and the potentials --- the RPA correlations included in their calculation which lead to a reduction of the cross section and a spreading of the spectrum.
To model the rescattering of the primary nucleons in the nucleus, Nieves \emph{et al.}~use a Monte Carlo simulation with elastic $NN$ cross section similar to ours.\footnote{We emphasize that, in addition, we allow for inelastic $NN$ collisions.} Therefore we expect a similar behavior when FSI are included. Indeed, as one can see, when the rescattering of the outgoing nucleons is "turned on", both calculations lead to very similar results, namely a reduction of the flux for higher energetic nucleons and a large number of secondary low energy nucleons (in \reffig{fig:nieves_comp} solid versus dash-dotted line). Also for the $p/n$ ratios we find reasonable agreement: our ratios plotted in \reffig{fig:influence_deltas_ratio} show a behavior similar to the ones of Nieves \emph{et al.}~(cf.~right panel of Fig.~17 in \refscite{Nieves:2005rq}).

Furthermore, we compare to the calculation of Martinez \refetal{Martinez:2005xe}. While our results \emph{without} FSI agree approximately with their relativistic plane wave impulse approximation (RPWIA) (up to small differences which can be attributed to the momentum distribution of the nucleons (relativistic mean-field versus local Fermi gas) and to the potentials used in the calculations), a comparison of the results \emph{with} FSI may be meaningless, because our method describes exclusive nucleon knockout reactions which may contain more than one nucleon in the outgoing channel and may leave the residual nucleus in a highly excited state. On the contrary, in their model, the FSI of the nucleon are considered within two frameworks. In one case, they use a relativistic optical potential in a distorted-wave impulse approximation (RDWIA), in the other case a relativistic multiple-scattering Glauber approximation (RMSGA) is used. These models can therefore account only for single-nucleon knockout. 
We emphasize that these absorption mechanisms as used by Martinez \emph{et al.}~can describe only the flux reduction at higher nucleon energies but not the rescattering in the medium which leads to the emission of a large number of lower energy secondary nucleons. Nucleons are not just absorbed but --- through rescattering --- ejected with a different energy, angle and/or charge. In addition, we stress that already at $E_{\nu} \approx 1$~GeV a large part of the ejected nucleons stems from $\Delta$ excitation (see \reffig{fig:nucl_sigmatot}), a process not contained in the calculation of \refcite{Martinez:2005xe}.

\section{Summary and conclusions}

In the present work we have investigated NC neutrino interactions with nucleons and nuclei extending our earlier work on CC reactions \cite{Leitner:2006ww}. 
The neutrino-nucleon reaction is dominated by quasielastic scattering and $\Delta$ excitation; we have found that the higher resonances only give a minor contribution at neutrino energies below $1.5$~GeV.  

Nuclear effects are included in the framework of a coupled-channel BUU transport theory (GiBUU model) where we account for in-medium modifications from Fermi motion, Pauli blocking, nuclear binding and collisional broadening of resonances. In this model, FSI are implemented by means of transport theory which permits exclusive reactions to be studied.  Within the same model, we can describe neutrino-induced pion production and nucleon knockout. In this respect, the model presented here is unique.

The pion production cross section is especially influenced by final-state interactions. In the elementary neutrino-nucleon reaction, more $\pi^0$ than charged pions are produced due to the isospin structure of resonance decay. When final-state interactions are included, those $\pi^0$ get absorbed or reinteract leading to side-feeding in the smaller $\pi^{\pm}$ channels. Quasielastic scattering followed by $\pi$ production in $NN$ collisions also accounts for a small fraction of the pion production cross section. 

Also for nucleon knockout the influence of the final-state interactions is significant. The rescattering of high energy nucleons in the nucleus leads to a reduction of higher kinetic energy nucleons while a large number of secondary nucleons at lower nucleon kinetic energies are ejected. Also in the case of nucleon knockout, we found that side-feeding is important. Furthermore, we have illustrated that for neutrino energies $\gtrsim 1$~GeV initial resonance excitation (predominantly $\Delta$) leads to a significant contribution to nucleon knockout.

Summarizing, we have found that in-medium modifications, and especially final-state interactions, have a big influence on the neutrino-nucleus cross sections. We emphasize that a good and well tested description of these effects is crucial for the understanding of current and future experiments.

\begin{acknowledgments}
This work has been supported by the Deutsche Forschungsgemeinschaft. 
\end{acknowledgments}



\begin{thebibliography}{10}

\bibitem{Ahrens:1986xe}
L.~A. Ahrens {\em et~al.},
\newblock Phys. Rev. {\bf D35}, 785 (1987).

\bibitem{Barish:1974fe}
S.~J. Barish {\em et~al.},
\newblock Phys. Rev. Lett. {\bf 33}, 448 (1974).

\bibitem{Derrick:1980nr}
M.~Derrick {\em et~al.},
\newblock Phys. Lett. {\bf B92}, 363 (1980).

\bibitem{Krenz:1977sw}
W.~Krenz {\em et~al.},
\newblock Nucl. Phys. {\bf B135}, 45 (1978).

\bibitem{Pohl:1978iy}
M.~Pohl {\em et~al.},
\newblock Phys. Lett. {\bf B79}, 501 (1978).

\bibitem{Nakayama:2004dp}
K2K, S.~Nakayama {\em et~al.},
\newblock Phys. Lett. {\bf B619}, 255 (2005), [hep-ex/0408134].

\bibitem{Raaf:2004ty}
BooNE, J.~L. Raaf,
\newblock Nucl. Phys. Proc. Suppl. {\bf 139}, 47 (2005), [hep-ex/0408015].

\bibitem{Harris:2004iq}
MINERvA, D.~A. Harris {\em et~al.},
\newblock hep-ex/0410005.

\bibitem{Fogli:1979qj}
G.~L. Fogli and G.~Nardulli,
\newblock Nucl. Phys. {\bf B165}, 162 (1980).

\bibitem{Rein:1980wg}
D.~Rein and L.~M. Sehgal,
\newblock Ann. Phys. {\bf 133}, 79 (1981).

\bibitem{Paschos:2000be}
E.~A. Paschos, L.~Pasquali and J.-Y. Yu,
\newblock Nucl. Phys. {\bf B588}, 263 (2000), [hep-ph/0005255].

\bibitem{Garvey:1992qp}
G.~T. Garvey, S.~Krewald, E.~Kolbe and K.~Langanke,
\newblock Phys. Lett. {\bf B289}, 249 (1992).

\bibitem{Horowitz:1993rj}
C.~J. Horowitz, H.-C. Kim, D.~P. Murdock and S.~Pollock,
\newblock Phys. Rev. {\bf C48}, 3078 (1993).

\bibitem{Barbaro:1996vd}
M.~B. Barbaro, A.~De~Pace, T.~W. Donnelly, A.~Molinari and M.~J. Musolf,
\newblock Phys. Rev. {\bf C54}, 1954 (1996), [nucl-th/9605020].

\bibitem{Alberico:1997vh}
W.~M. Alberico, M.~B. Barbaro, S.~M. Bilenky, J.~A. Caballero, C.~Giunti,
  C.~Maieron, E.~Moya~de Guerra and J.~M. Udias,
\newblock Nucl. Phys. {\bf A623}, 471 (1997), [hep-ph/9703415].

\bibitem{Alberico:1997rm}
W.~M. Alberico, M.~B. Barbaro, S.~M. Bilenky, J.~A. Caballero, C.~Giunti,
  C.~Maieron, E.~Moya~de Guerra and J.~M. Udias,
\newblock Phys. Lett. {\bf B438}, 9 (1998), [hep-ph/9712441].

\bibitem{Meucci:2004ip}
A.~Meucci, C.~Giusti and F.~D. Pacati,
\newblock Nucl. Phys. {\bf A744}, 307 (2004), [nucl-th/0405004].

\bibitem{Martinez:2005xe}
M.~C. Martinez, P.~Lava, N.~Jachowicz, J.~Ryckebusch, K.~Vantournhout and J.~M.
  Udias,
\newblock Phys. Rev. {\bf C73}, 024607 (2006), [nucl-th/0505008].

\bibitem{vanderVentel:2005ke}
B.~I.~S. van~der Ventel and J.~Piekarewicz,
\newblock Phys. Rev. {\bf C73}, 025501 (2006), [nucl-th/0506071].

\bibitem{Meucci:2006ir}
A.~Meucci, C.~Giusti and F.~D. Pacati,
\newblock Nucl. Phys. {\bf A773}, 250 (2006), [nucl-th/0601052].

\bibitem{Kolbe:1994xb}
E.~Kolbe, K.~Langanke and S.~Krewald,
\newblock Phys. Rev. {\bf C49}, 1122 (1994).

\bibitem{Jachowicz:1998fn}
N.~Jachowicz, S.~Rombouts, K.~Heyde and J.~Ryckebusch,
\newblock Phys. Rev. {\bf C59}, 3246 (1999).

\bibitem{Umino:1994wu}
Y.~Umino, J.~M. Udias and P.~J. Mulders,
\newblock Phys. Rev. Lett. {\bf 74}, 4993 (1995).

\bibitem{Amaro:2006pr}
J.~E. Amaro, M.~B. Barbaro, J.~A. Caballero and T.~W. Donnelly,
\newblock Phys. Rev. {\bf C73}, 035503 (2006), [nucl-th/0602053].

\bibitem{Nieves:2005rq}
J.~Nieves, M.~Valverde and M.~J. Vicente~Vacas,
\newblock Phys. Rev. {\bf C73}, 025504 (2006), [hep-ph/0511204].

\bibitem{Leitner:2006ww}
T.~Leitner, L.~Alvarez-Ruso and U.~Mosel,
\newblock Phys. Rev. {\bf C73}, 065502 (2006), [nucl-th/0601103].

\bibitem{Casper:2002sd}
D.~Casper,
\newblock Nucl. Phys. Proc. Suppl. {\bf 112}, 161 (2002), [hep-ph/0208030].

\bibitem{Gallagher:2002sf}
H.~Gallagher,
\newblock Nucl. Phys. Proc. Suppl. {\bf 112}, 188 (2002).

\bibitem{Hayato:2002sd}
Y.~Hayato,
\newblock Nucl. Phys. Proc. Suppl. {\bf 112}, 171 (2002).

\bibitem{Tiator:2003uu}
L.~Tiator, D.~Drechsel, S.~Kamalov, M.~M. Giannini, E.~Santopinto and
  A.~Vassallo,
\newblock Eur. Phys. J. {\bf A19}, 55 (2004), [nucl-th/0310041].

\bibitem{Lalakulich:2006sw}
O.~Lalakulich, E.~A. Paschos and G.~Piranishvili,
\newblock Phys. Rev. {\bf D74}, 014009 (2006), [hep-ph/0602210].

\bibitem{Budd:2003wb}
H.~Budd, A.~Bodek and J.~Arrington,
\newblock hep-ex/0308005.

\bibitem{Bernard:2001rs}
V.~Bernard, L.~Elouadrhiri and U.~G. Meissner,
\newblock J. Phys. {\bf G28}, R1 (2002), [hep-ph/0107088].

\bibitem{Weise:2001}
A.~W. Thomas and W.~Weise,
\newblock {\em The Structure of the Nucleon} (Wiley-VCH, Berlin, 2001).

\bibitem{Spayde:2003nr}
SAMPLE, D.~T. Spayde {\em et~al.},
\newblock Phys. Lett. {\bf B583}, 79 (2004), [nucl-ex/0312016].

\bibitem{Aniol:2000at}
HAPPEX, K.~A. Aniol {\em et~al.},
\newblock Phys. Lett. {\bf B509}, 211 (2001), [nucl-ex/0006002].

\bibitem{Aniol:2005zg}
HAPPEX, K.~A. Aniol {\em et~al.},
\newblock Phys. Lett. {\bf B635}, 275 (2006), [nucl-ex/0506011].

\bibitem{Armstrong:2005hs}
G0, D.~S. Armstrong {\em et~al.},
\newblock Phys. Rev. Lett. {\bf 95}, 092001 (2005), [nucl-ex/0506021].

\bibitem{Maas:2004ta}
A4, F.~E. Maas {\em et~al.},
\newblock Phys. Rev. Lett. {\bf 93}, 022002 (2004), [nucl-ex/0401019].

\bibitem{Young:2006jc}
R.~D. Young, J.~Roche, R.~D. Carlini and A.~W. Thomas,
\newblock Phys. Rev. Lett. {\bf 97}, 102002 (2006), [nucl-ex/0604010].

\bibitem{Garvey:1992cg}
G.~T. Garvey, W.~C. Louis and D.~H. White,
\newblock Phys. Rev. {\bf C48}, 761 (1993).

\bibitem{Alberico:1998qw}
W.~M. Alberico, M.~B. Barbaro, S.~M. Bilenky, J.~A. Caballero, C.~Giunti,
  C.~Maieron, E.~Moya~de Guerra and J.~M. Udias,
\newblock Nucl. Phys. {\bf A651}, 277 (1999), [hep-ph/9812388].

\bibitem{Pate:2003rk}
S.~F. Pate,
\newblock Phys. Rev. Lett. {\bf 92}, 082002 (2004), [hep-ex/0310052].

\bibitem{Pate:2005ft}
S.~F. Pate,
\newblock Eur. Phys. J. {\bf A24S2}, 67 (2005), [nucl-ex/0502014].

\bibitem{Pate:2005bk}
S.~F. Pate, G.~A. MacLachlan, D.~W. McKee and V.~Papavassiliou,
\newblock AIP Conf. Proc. {\bf 842}, 309 (2006), [hep-ex/0512032].

\bibitem{finesseprop}
FINeSSE, L.~Bugel {\em et~al.},
\newblock hep-ex/0402007.

\bibitem{leitner_diplom2}
T.~Leitner,
\newblock {\em Neutrino Interactions with Nucleons and Nuclei},
\newblock Diploma thesis, Universit\"at Giessen, 2005,
\newblock available online at
  http://theorie.physik.uni-giessen.de/documents/diplom/leitner.pdf.

\bibitem{LlewellynSmith:1971zm}
C.~H. Llewellyn~Smith,
\newblock Phys. Rept. {\bf 3}, 261 (1972).

\bibitem{Paschos:2003qr}
E.~A. Paschos, J.-Y. Yu and M.~Sakuda,
\newblock Phys. Rev. {\bf D69}, 014013 (2004), [hep-ph/0308130].

\bibitem{Fogli:1979cz}
G.~L. Fogli and G.~Nardulli,
\newblock Nucl. Phys. {\bf B160}, 116 (1979).

\bibitem{Alvarez-Ruso:1997jr}
L.~Alvarez-Ruso, S.~K. Singh and M.~J. Vicente~Vacas,
\newblock Phys. Rev. {\bf C57}, 2693 (1998), [nucl-th/9712058].

\bibitem{Alvarez-Ruso:2003gj}
L.~Alvarez-Ruso, M.~B. Barbaro, T.~W. Donnelly and A.~Molinari,
\newblock Nucl. Phys. {\bf A724}, 157 (2003), [nucl-th/0303027].

\bibitem{Li:1991yb}
Z.-p. Li, V.~Burkert and Z.-j. Li,
\newblock Phys. Rev. {\bf D46}, 70 (1992).

\bibitem{Cano:1998wz}
F.~Cano and P.~Gonzalez,
\newblock Phys. Lett. {\bf B431}, 270 (1998), [nucl-th/9804071].

\bibitem{Alberto:2001fy}
P.~Alberto, M.~Fiolhais, B.~Golli and J.~Marques,
\newblock Phys. Lett. {\bf B523}, 273 (2001), [hep-ph/0103171].

\bibitem{Cardarelli:1996vn}
F.~Cardarelli, E.~Pace, G.~Salme and S.~Simula,
\newblock Phys. Lett. {\bf B397}, 13 (1997), [nucl-th/9609047].

\bibitem{Dong:1999cz}
Y.~B. Dong, K.~Shimizu, A.~Faessler and A.~J. Buchmann,
\newblock Phys. Rev. {\bf C60}, 035203 (1999).

\bibitem{Zeller:2003ey}
G.~P. Zeller,
\newblock hep-ex/0312061.

\bibitem{manley}
D.~M. Manley and E.~M. Saleski,
\newblock Phys. Rev. {\bf D45}, 4002 (1992).

\bibitem{hawker}
E.~A. Hawker,
\newblock (2002),
\newblock available online at
  http://www.ps.uci.edu/nuint/proceedings/hawker.pdf.

\bibitem{Nieves:1993ev}
J.~Nieves, E.~Oset and C.~Garcia-Recio,
\newblock Nucl. Phys. {\bf A554}, 509 (1993).

\bibitem{Effenberger:1999ay}
M.~Effenberger, E.~L. Bratkovskaya and U.~Mosel,
\newblock Phys. Rev. {\bf C60}, 044614 (1999), [nucl-th/9903026].

\bibitem{Ericson:1988gk}
T.~E.~O. Ericson and W.~Weise,
\newblock {\em Pions and Nuclei} (Clarendon Press, Oxford, 1988).

\bibitem{gibuu}
Gi{BUU} website,
\newblock http://theorie.physik.uni-giessen.de/GiBUU.

\bibitem{Teis:1996kx}
S.~Teis, W.~Cassing, M.~Effenberger, A.~Hombach, U.~Mosel and G.~Wolf,
\newblock Z. Phys. {\bf A356}, 421 (1997), [nucl-th/9609009].

\bibitem{Hombach:1998wr}
A.~Hombach, W.~Cassing, S.~Teis and U.~Mosel,
\newblock Eur. Phys. J. {\bf A5}, 157 (1999), [nucl-th/9812050].

\bibitem{Wagner:2004ee}
M.~Wagner, A.~B. Larionov and U.~Mosel,
\newblock Phys. Rev. {\bf C71}, 034910 (2005), [nucl-th/0411010].

\bibitem{Weidmann:1997vj}
T.~Weidmann, E.~L. Bratkovskaya, W.~Cassing and U.~Mosel,
\newblock Phys. Rev. {\bf C59}, 919 (1999), [nucl-th/9711004].

\bibitem{Lehr:1999zr}
J.~Lehr, M.~Effenberger and U.~Mosel,
\newblock Nucl. Phys. {\bf A671}, 503 (2000), [nucl-th/9907091].

\bibitem{Falter:2004uc}
T.~Falter, W.~Cassing, K.~Gallmeister and U.~Mosel,
\newblock Phys. Rev. {\bf C70}, 054609 (2004), [nucl-th/0406023].

\bibitem{Buss:2006vh}
O.~Buss, L.~Alvarez-Ruso, P.~Muehlich and U.~Mosel,
\newblock Eur. Phys. J. {\bf A29 (2)}, 189 (2006), [nucl-th/0603003].

\bibitem{Buss:2006yk}
O.~Buss, L.~Alvarez-Ruso, A.~B. Larionov and U.~Mosel,
\newblock Phys. Rev. {\bf C74}, 044610 (2006), [nucl-th/0607016].

\bibitem{Cugnon:1996kh}
J.~Cugnon, J.~Vandermeulen and D.~L'Hote,
\newblock Nucl. Instrum. Meth. {\bf B111}, 215 (1996).

\bibitem{Oset:1987re}
E.~Oset and L.~L. Salcedo,
\newblock Nucl. Phys. {\bf A468}, 631 (1987).

\bibitem{Engel:1993jh}
A.~Engel, W.~Cassing, U.~Mosel, M.~Schafer and G.~Wolf,
\newblock Nucl. Phys. {\bf A572}, 657 (1994), [nucl-th/9307008].

\bibitem{Falter:2003uy}
T.~Falter, J.~Lehr, U.~Mosel, P.~Muehlich and M.~Post,
\newblock Prog. Part. Nucl. Phys. {\bf 53}, 25 (2004), [nucl-th/0312093].

\bibitem{Alvarez-Ruso:2004ji}
L.~Alvarez-Ruso, T.~Falter, U.~Mosel and P.~Muehlich,
\newblock Prog. Part. Nucl. Phys. {\bf 55}, 71 (2005), [nucl-th/0412084].

\bibitem{Adler:1974qu}
S.~L. Adler, S.~Nussinov and E.~A. Paschos,
\newblock Phys. Rev. {\bf D9}, 2125 (1974).

\bibitem{Paschos:2003ej}
E.~A. Paschos, D.~P. Roy, I.~Schienbein and J.~Y. Yu,
\newblock Phys. Lett. {\bf B574}, 232 (2003), [hep-ph/0307223].

\bibitem{Paschos:2004qh}
E.~A. Paschos, I.~Schienbein and J.~Y. Yu,
\newblock Nucl. Phys. Proc. Suppl. {\bf 139}, 119 (2005), [hep-ph/0408148].

\bibitem{Krusche:2004uw}
B.~Krusche {\em et~al.},
\newblock Eur. Phys. J. {\bf A22}, 277 (2004), [nucl-ex/0406002].

\bibitem{paschosprivcomm}
E.~A. Paschos,
\newblock private communication (September 2006).

\end{thebibliography}
\end{document}